\documentstyle[12pt,epsf]{article}

\newcommand{\beq}{\begin{equation}}
\newcommand{\eeq}{\end{equation}}
\newcommand{\bea}{\begin{eqnarray}}
\newcommand{\eea}{\end{eqnarray}}
\newcommand{\rar}{\rightarrow}

\newcommand{\lan}{\langle}
\newcommand{\ran}{\rangle}
\newcommand{\balpha}{\mbox{\boldmath$\alpha$}}

\begin{document}

\font\fortssbx=cmssbx10 scaled \magstep2
\hbox to \hsize{
\includegraphics{uwlogo.ps}
\hskip.5in \raise.1in\hbox{\fortssbx University of Wisconsin - Madison}
\hfill$\vcenter{\hbox{\bf MADPH-95-907}
            \hbox{September 1995}}$ }
\vskip 2cm
\begin{center}
\Large
{\bf $S$ to $P$ wave
form factors in semi-leptonic $B$ decays} \\
\vskip 0.5cm
\large
  Sini\v{s}a Veseli  and M. G. Olsson  \\
\vskip 0.1cm
{\small \em Department of Physics, University of Wisconsin, Madison,
	\rm WI 53706}
\end{center}
\thispagestyle{empty}
\vskip 0.7cm

\begin{abstract}
We apply HQET to semi-leptonic $B$ and $B_{s}$ meson decays into
the observed charmed $P$ wave states. In order to examine the sensitivity
of the results to the choice of a specific model, we perform all
calculations using several different meson models, and find that
uncertainty introduced by the choice of a particular model is about 30\%.
Specifically, assuming $\tau_{B}=1.50ps$ and $V_{cb}= 0.040$, we obtain
branching ratios of $(0.27\pm 0.08)\%$ and $(0.45\pm 0.14)\%$ for
$B\rar D_{1}l\bar{\nu}_{l}$ and $B\rar D_{2}^{*}l\bar{\nu}_{l}$ decays,
respectively.
\end{abstract}

\newpage
\section{Introduction}

As more $B$ mesons are produced at major accelerators it
has become imperative to gain understanding of how they decay.
If the final hadron state consists of a meson containing
a heavy quark,  heavy quark symmetry \cite{isgur} and
heavy quark effective theory (HQET) \cite{georgi,reviews} provide a powerful
assistance.

Since an infinitely massive heavy quark does not
recoil from the emission or absorption of soft ($E\approx \Lambda_{QCD}$)
 gluons, and since magnetic interactions of such a
quark are negligible ($\sim \frac{1}{m_{Q}}$), the strong interactions
of the heavy quark are independent of its mass and spin,  and
the total angular momentum $j$
of the  LDF is a good quantum number.
Because of this,
HQET leads to relations between different
form factors describing
transitions in which a hadron containing a heavy quark $Q$ and
moving with four-velocity $v^{\mu}$, decays into another hadron
containing a heavy quark $Q'$, and moving with four-velocity $v'^{\mu}$.
In this way the number of independent form factors for these
decays is significantly reduced.

Semi-leptonic decays into hadrons account for over 20\% of
all $B$ decays. In the case of $B^{-}$ meson decaying
into electron, neutrino, and all hadrons the branching ratio is
\cite{gronberg}
\beq
BR(B^{-}\rar Xe^{-}\bar{\nu}) = (10.49\pm 0.46)\%\ .
\eeq
Most of the inclusive rate is accounted for
by $X=D$ and $X=D^{*}(2010)$. The measured branching ratios for
these final states are \cite{cleo}
\bea
BR(B^{-}\rar D^{*}e^{-}\bar{\nu}) &=& (5.13\pm 0.84)\%\ ,\\
BR(B^{-}\rar De^{-}\bar{\nu}) &=& (1.95\pm 0.55)\%\ .
\eea
This leaves $(3.4\pm1.1)\%$ of the hadrons unaccounted for.

In this paper we investigate $B$ decays
into the $P$ wave $D$ meson
states $D_{1}(2420)$ and $D_{2}^{*}(2460)$,
for which some experimental data is becoming available, and
also the corresponding $B_{s}$ decays.
We use the covariant trace formalism \cite{georgi2,korner,falk}
and HQET to obtain expressions for
branching ratios in terms of the non-perturbative Isgur-Wise (IW)
form factors. In order to calculate these form factors
we employ
expressions (consistent
with the trace formalism),  in terms of the light degrees
of freedom (LDF) wave functions and energies \cite{modelling}.
By performing all calculations using four different models
and two different one basis state estimates, we also examine
sensitivity of our results to the choice of a specific model.

In Sections \ref{trace}, \ref{rates} and \ref{iwf} we review
the covariant representation of states, decay rates for
$B\rar D_{X}l\bar{\nu}_{l}$, and the calculation of IW functions,
respectively. In Section \ref{models} we discuss the four
heavy-light models which are employed in this paper. We also
discuss estimates which use only one basis state as the wave
function of the LDF. The model dependence of our results
will be judged by the range of prediction of these calculations.
Our main results for the $S$ to $P$ wave
semi-leptonic branching ratios for $B$ decay into
$D_{1}$ and $D_{2}^{*}$ (and corresponding $B_{s}$ decays),
are given in Section \ref{results}. Our conclusions
and a comparison with experiment
are summarized
in Section \ref{con}.

\section{Covariant representation of states}
\label{trace}

The covariant trace formalism, formulated
in \cite{georgi2,korner} and generalized
to  excited states in \cite{falk},
is most convenient for  counting of the number of independent
form factors.
Following \cite{falk}, and
using the notation of \cite{mannel},
the lowest lying mesonic states with mass $m$ and
four velocity $v$ can be described as follows:
\bea
C(v)& =& \frac{1}{2}\sqrt{m} (\not{v}+1)\gamma_{5}\ ,
\hspace*{+4.35cm}\ J^{P}=0^{-}\ ,
\ j=\frac{1}{2}\ ,\label{st1}\\
C^{*}(v,\epsilon)& =& \frac{1}{2}\sqrt{m} (\not{v}+1)\not{\epsilon}
\ ,
\hspace*{+4.3cm}\ J^{P}=1^{-}\ ,
\ j=\frac{1}{2}\ , \\
E(v)& =& \frac{1}{2}\sqrt{m} (\not{v}+1)\ ,
\hspace*{+4.65cm}
\ J^{P}=0^{+}\ ,
\ j=\frac{1}{2}\ ,\label{st3}\\
E^{*}(v,\epsilon)& =& \frac{1}{2}\sqrt{m} (\not{v}+1)
\gamma_{5}\not{\epsilon}
\ ,
\hspace*{+3.85cm}\ J^{P}=1^{+}\ ,
\ j=\frac{1}{2}\ ,\\
F(v,\epsilon)& =& \frac{1}{2}\sqrt{m}
\sqrt{\frac{3}{2}} (\not{v}+1)\gamma_{5}[
\epsilon^{\mu}-\frac{1}{3}\not{\epsilon}(\gamma^{\mu}-v^{\mu})]
\ ,
\hspace*{+0.1cm}
\ J^{P}=1^{+}\ ,
\ j=\frac{3}{2}\ ,\label{st5}\\
F^{*}(v,\epsilon)& =& \frac{1}{2}\sqrt{m} (\not{v}+1)
\gamma_{\nu}\epsilon^{\mu\nu}
\ ,
\hspace*{+3.65cm}
\ J^{P}=2^{+}\ ,
\ j=\frac{3}{2}\ ,\\
G(v,\epsilon)& =& \frac{1}{2}\sqrt{m}
\sqrt{\frac{3}{2}} (\not{v}+1)[
\epsilon^{\mu}-\frac{1}{3}\not{\epsilon}(\gamma^{\mu}+v^{\mu})]
\ ,
\hspace*{+0.6cm}
\ J^{P}=1^{-}\ ,
\ j=\frac{3}{2}\ ,\\
G^{*}(v,\epsilon)& =& \frac{1}{2}\sqrt{m} (\not{v}+1)
\gamma_{5}\gamma_{\nu}\epsilon^{\mu\nu}
\ ,
\hspace*{+3.4cm}\ J^{P}=2^{-}\ ,
\ j=\frac{3}{2}\ .\label{st8}
\eea
In these expressions
 $\epsilon^{\mu}$ is the polarization vector for spin $1$ states
(satisfying $\epsilon\cdot v=0$), while the tensor $\epsilon^{\mu\nu}$
describes a spin 2 object ($\epsilon^{\mu\nu}=\epsilon^{\nu\mu}\ ,\
\epsilon^{\mu\nu}v_{\nu}=0\ ,\ \epsilon^{\mu}_{\ \mu}=0$).
For each $j$ there are two
degenerate heavy meson states ($J=j\pm \frac{1}{2}$)
 forming a spin symmetry doublet:
 ($C$, $C^{*})$ is the
 $L=0$ doublet, ($E$, $E^{*}$) and  ($F$, $F^{*}$) are the two
$L=1$ doublets, and ($G$, $G^{*}$) is an $L=2$ doublet.

In the covariant trace formalism
matrix elements of bilinear  currents of
two heavy quarks ($J(q)=\bar{Q'}\Gamma Q$)
between the physical meson states are calculated
by taking the trace ($\omega=v\cdot v'$),
\beq
\lan \Psi'(v')|J(q)|\Psi(v)\ran =  {\rm Tr}[\bar{M'}(v')\Gamma M(v)]
{\cal M}_{l}(\omega)\ ,\label{tr}
\eeq
where $M'$ and $M$ denote appropriate matrices from
(\ref{st1})-(\ref{st8}), $\bar{M}=\gamma^{0}M^{\dag}\gamma^{0}$, and
${\cal M}_{l}(\omega)$ represents the LDF.
Again following \cite{falk,mannel}, we define the IW functions
for the transitions of a $0^{-}$ ground state into an excited state by
\beq
{\cal M}_{l}(\omega)
=
\left\{
\begin{array}{ll}
 \xi_{C}(\omega)\ , &
C\rar(C,C^{*})\ , \\
 \xi_{E}(\omega)\ , &
 C\rar(E,E^{*})\ ,\\
\xi_{F}(\omega)v_{\mu}\ ,&
C\rar(F,F^{*})\ ,\\
\xi_{G}(\omega)v_{\mu}\ ,&
C\rar(G,G^{*})\ .
\end{array}\right.
\label{xie}
\eeq
The vector index in the last two definitions will be contracted
with the one
in the representations of excited states (\ref{st5})-(\ref{st8}).

\section{Decays $B\rar D_{X} l\bar{\nu}_{l}$ in the
heavy quark limit}
\label{rates}

Denoting the four-velocities of $B$ and $D_{X}$ mesons as
$v^{\mu}$ and $v'^{\mu}$, respectively, and assuming that
lepton   masses are zero, the momentum transfer
is given by ($\omega=v\cdot v'$)
\beq
q^{2} = (m_{B}v-m_{D_{X}}v')^{2} = (p_{1}+p_{2})^{2}
= m_{B}^{2}+m_{D_{X}}^{2} - 2m_{B}m_{D_{X}}\omega\ .
\label{q2}
\eeq
Using (\ref{q2}), and denoting
\beq
x = (p_{1}+m_{D_{X}}v')^{2} = (m_{B}v-p_{2})^{2}\ ,
\label{xdef}
\eeq
the standard expression \cite{pdg} for the width of the semi-leptonic
decay
of a $B$ meson into  any of the charmed meson states $D_{X}$
can  be written as
\beq
\frac{d\Gamma}{d\omega}=
\frac{ m_{D_{X}}}{128\pi^{3}m_{B}^{2}}
 \int_{x_{-}}^{x_{+}}dx \overline{|{\cal M}|^{2}}\ ,
\label{dg}
\eeq
where
$x_{\pm} = m_{B}m_{D_{X}}(\omega \pm \sqrt{\omega^{2}-1})$.
The invariant amplitude $\cal M$,
\beq
{\cal M} = \frac{G_{F}V_{cb}}{\sqrt{2}}\bar{u}_{l}
\gamma^{\mu}(1-\gamma^{5})v_{\bar{\nu}}
\lan D_{X}(v',\epsilon')|\bar{c}\gamma_{\mu}(1-\gamma^{5})b|B(v)\ran\ ,
\label{m}
\eeq
after squaring and summing over $l$ and $\bar{\nu}_{l}$
spins and
$D_{X}$ polarization,
yields
\beq
 \overline{|{\cal M}|^{2}} =
\frac{1}{2}G_{F}^{2}|V_{cb}|^{2}L_{\mu\nu}H^{\mu\nu}\ .
\eeq
Here,
\bea
L^{\mu\nu}& =& 8(p_{1}^{\mu}p_{2}^{\nu}-g^{\mu\nu}p_{1}\cdot p_{2}
+p_{1}^{\nu}p_{2}^{\mu}+i\varepsilon^{\mu\nu\alpha\beta}p_{1\alpha}
p_{2\beta})\ ,\\
H^{\mu\nu} &=& \sum_{pol}
\lan D_{X}(v',\epsilon')|\bar{c}\gamma^{\mu}(1-\gamma^{5})b|B(v)\ran
\lan D_{X}(v',\epsilon')|\bar{c}
\gamma^{\nu}(1-\gamma^{5})b|B(v)\ran^{\dagger}\ .
\eea
The matrix elements needed in $H^{\mu\nu}$ are calculated from
(\ref{tr}) using  (\ref{st1})-(\ref{st8}), while the
sum over $D_{X}$ polarization states
is performed using standard expressions for spin-1 and spin-2
particles,
\bea
M_{\mu\nu}^{(1)}(v) &\equiv&\sum_{pol}\epsilon^{*}_{\mu}\epsilon_{\nu}
\nonumber \\
&=& -g_{\mu\nu} + v_{\mu}v_{\nu}\ ,\\
M_{\mu\nu,\rho\sigma}^{(2)}(v)
&\equiv&\sum_{pol}\epsilon^{*}_{\mu\nu}\epsilon_{\rho\sigma}
\nonumber \\
&=&\frac{1}{2}M_{\mu\rho}^{(1)}(v)M_{\nu\sigma}^{(1)}(v)
+\frac{1}{2}M_{\mu\sigma}^{(1)}(v)M_{\nu\rho}^{(1)}(v)
-\frac{1}{3}M_{\mu\nu}^{(1)}(v)M_{\rho\sigma}^{(1)}(v)\ .
\eea
Using kinematical identities
coming from definitions (\ref{q2}) and (\ref{xdef}),
and from momentum conservation, we can express
$\overline{|{\cal M}|^{2}}$ in terms of $\omega$ and $x$. Performing
a simple integration in (\ref{dg}), we find
\beq
\frac{d\Gamma_{X}}{d\omega} = \frac{G_{F}^{2}|V_{cb}|^{2}}{48\pi^{3}}
m_{B}^{2}m_{D_{X}}^{3}\sqrt{\omega^{2}-1}|\xi_{X}(\omega)|^{2}
f_{X}(\omega,r_{X})\ ,
\label{width}
\eeq
where $r_{X}=m_{D_{X}}/m_{B}$, and the function $f_{X}$
is given by
\bea
f_{C}(\omega,r_{C}) &\hspace*{-2mm}=\hspace*{-2mm}&
(\omega^{2}-1)(1+r_{C})^{2}\ ,\\
f_{C^{*}}(\omega,r_{C^{*}}) &\hspace*{-2mm}=\hspace*{-2mm}&
(\omega+1)[(\omega+1)(1-r_{C^{*}})^{2} +
4\omega(1-2\omega r_{C^{*}} + r_{C^{*}}^{2})]\ ,\\
f_{ E}(\omega,r_{ E}) &\hspace*{-2mm}=\hspace*{-2mm}&
(\omega^{2}-1)(1-r_{ E})^{2}\ ,\\
f_{E^{*}}(\omega,r_{E^{*}}) &\hspace*{-2mm}=\hspace*{-2mm}&
(\omega-1)[(\omega-1)(1+r_{E^{*}})^{2} +
4\omega(1-2\omega r_{E^{*}} + r_{E^{*}}^{2})]\ ,\\
f_{F}(\omega,r_{F}) &\hspace*{-2mm}=\hspace*{-2mm}&
\frac{2}{3}(\omega-1)(\omega+1)^{2}
[(\omega-1)(1+r_{F})^{2} +
\omega(1-2\omega r_{F} + r_{F}^{2})]\ ,\\
f_{F^{*}}(\omega,r_{F^{*}}) &\hspace*{-2mm}=\hspace*{-2mm}&
\frac{2}{3}(\omega-1)(\omega+1)^{2}
[(\omega+1)(1-r_{F^{*}})^{2} +
3\omega(1-2\omega r_{F^{*}} + r_{F^{*}}^{2})]\ ,\\
f_{G}(\omega,r_{G}) &\hspace*{-2mm}=\hspace*{-2mm}&
\frac{2}{3}(\omega-1)^{2}(\omega+1)
[(\omega+1)(1-r_{G})^{2} +
\omega(1-2\omega r_{G} + r_{G}^{2})]\ ,\\
f_{G^{*}}(\omega,r_{G^{*}}) &\hspace*{-2mm}=\hspace*{-2mm}&
\frac{2}{3}(\omega-1)^{2}(\omega+1)
[(\omega-1)(1+r_{G^{*}})^{2} +
3\omega(1-2\omega r_{G^{*}} + r_{G^{*}}^{2})]\ .
\eea
Some, but not all of the above expressions can be found in earlier work
\cite{neubert}-\cite{suzuki}.

\section{IW functions}
\label{iwf}

The only  factor
in (\ref{width})
which
cannot be calculated from   first principles
is the IW function
for a particular decay, .
In order to estimate these form factors
one  has to rely on some model of strong interactions.
In the original calculation of
radiative rare $B$ decays
\cite{mannel} the IW functions (\ref{xie}) were defined
 as
the overlap between wave functions describing
 the LDF in the initial and the
final mesons (AOM).
These authors have chosen the wave functions
to be eigenfunctions of orbital angular momentum
$L$ ($\alpha$ denotes all other quantum numbers),
\beq
\Phi_{\alpha Lm_{L}}({\bf x}) = R_{\alpha L}(r) Y_{Lm_{L}}(\Omega)\ ,
\eeq
and the form factors were given by
(putting a
tilde  to avoid confusion with our definitions),
\bea
\tilde{\xi}_{C}(\omega)& =& \lan j_{0}(\tilde{a}r)\ran_{00}\ ,
\label{xicmannel}
\\
\tilde{\xi}_{E}(\omega)& =& \sqrt{3}\lan j_{1}(\tilde{a}r)\ran_{10}\ ,
\\
\tilde{\xi}_{F}(\omega)& =& \sqrt{3}\lan j_{1}(\tilde{a}r)\ran_{10}\ ,
\label{xifmannel}
\\
\tilde{\xi}_{G}(\omega)& =& \sqrt{5}\lan j_{2}(\tilde{a}r)\ran_{20}\ .
\label{xigmannel}
\eea
In the above
\beq
\tilde{a}=\tilde{E}'_{\bar{q}}\sqrt{\omega^{2}-1}\ ,
\eeq
where $\tilde{E}'_{\bar{q}} =
\frac{M' m_{\bar{q}}}{m_{Q'}+m_{\bar{q}}}$
denotes the ``inertia parameter'' of the LDF in the final
heavy meson (with
 mass M' and heavy quark $Q'$). The expectation values in
(\ref{xicmannel})-(\ref{xigmannel}) are defined by
\beq
\lan F(r) \ran_{L'L}^{\alpha'\alpha} = \int r^{2} dr R^{*}_{\alpha'L'}(r)
R_{\alpha L}(r)F(r)\ .
\label{jave}
\eeq
To calculate the above overlap integrals, in \cite{mannel} the radial
wave functions of the ISGW model \cite{isgw} were used. The same
approach was followed in \cite{suzuki} for the calculation
of semi-leptonic $B$  meson decays into higher charmed resonances.
However, by comparing the covariant trace formalism of
\cite{georgi2,korner,falk} with the wave function approach
of \cite{zalewski2}, it has been
 recently shown \cite{modelling} that
form factor definitions as pure overlap integrals
(used in \cite{mannel,suzuki}), are
not consistent with the  trace formalism.

Under
the assumption that heavy mesons can be described using simple
non-relativistic (or semi-relativistic) quark model,
the rest frame LDF wave functions
(with angular momentum $j$ and its projection $\lambda_{j}$),
can be written as
\beq
\phi^{(\alpha L)}_{j\lambda_{j}}({\bf x})= \sum_{m_{L},m_{s}}
R_{\alpha L}(r)Y_{Lm_{L}}(\Omega)\chi_{m_{s}}
\lan L,m_{L};\frac{1}{2},m_{s}|j,\lambda_{j};L,\frac{1}{2}\ran\ ,
\label{wf}
\eeq
where $\chi_{m_{s}}$ represent the rest frame spinors normalized
to one, $\chi^{\dag}_{m'_{s}}\chi_{m_{s}}=\delta_{m'_{s},m_{s}}$, and
$\alpha$ again represents all other quantum numbers.
Performing the overlap integrals in the modified Breit
frame (${\bf v}' = -{\bf v})$ \cite{zalewski},  using (\ref{wf}) and
form factor definitions
 consistent with the  trace formalism,   one can derive
\cite{modelling}  expressions
for the IW functions, their
 values,
and values of their derivatives at the zero recoil point ($\omega = 1$),
 in terms of
the LDF wave functions.
Denoting ($E_{\bar{q}} = M - m_{Q}$ here refers to the  LDF energy of
a meson with mass $M$ and heavy quark $Q$),
\beq
a = (E_{\bar{q}}+E'_{\bar{q}})\sqrt{\frac{\omega-1}{\omega+1}}\ ,
\eeq
and
suppressing
quantum numbers $\alpha'$ and $\alpha$, we have:
\begin{itemize}
\item $C\rar (C,C^{*})$ transitions.
\bea
\xi_{C}(\omega)& =&
\frac{2}{\omega+1}
\lan j_{0}(ar)\ran_{00} \ ,\label{xicq}\\
\xi_{C}(1)& =& \lan 1\ran_{00}\ ,\\
\xi'_{C}(1) &=& -\frac{1}{2} - \frac{1}{12}
(E_{\bar{q}}+ E'_{\bar{q}})^{2}<r^{2}>_{00}\ .\label{xicsl}
\eea
\item $C\rar (E,E^{*})$ transitions.
\bea
\xi_{E}(\omega)& =&
\frac{2}{\sqrt{\omega^{2}-1}}
\lan j_{1}(ar)\ran_{10}\label{xieq}
 \ ,\\
\xi_{E}(1)& =& \frac{1}{3}(E_{\bar{q}}+E'_{\bar{q}})\lan r\ran_{10}\ ,\\
\xi'_{E}(1) &=& -\frac{1}{6}(E_{\bar{q}}+E'_{\bar{q}})\lan r\ran_{10}
 - \frac{1}{60} (E_{\bar{q}}+E'_{\bar{q}})^{3}<r^{3}>_{10}\ .
\label{dxieq1}
\eea
\item $C\rar (F,F^{*})$ transitions.
\bea
\xi_{F}(\omega)& =&
\sqrt{\frac{3}{\omega^{2}-1}}\frac{2}{\omega+1}
\lan j_{1}(ar)\ran_{10} \label{xifq}\ ,\\
\xi_{F}(1)& =& \frac{1}{2\sqrt{3}}
(E_{\bar{q}}+E'_{\bar{q}})\lan r\ran_{10}\ ,
\label{xifq1}\\
\xi'_{F}(1) &=& -\frac{1}{2\sqrt{3}}(E_{\bar{q}}+E'_{\bar{q}})\lan r\ran_{10}
 - \frac{1}{40\sqrt{3}} (E_{\bar{q}}+E'_{\bar{q}})^{3}<r^{3}>_{10}\ .
\label{dxifq1}
\eea
\item $C\rar (G,G^{*})$ transitions.
\bea
\xi_{G}(\omega)& =&
\frac{2\sqrt{3}}{\omega^{2}-1}
\lan j_{2}(ar)\ran_{20} \ ,\\
\xi_{G}(1)& =& \frac{1}{10\sqrt{3}}
(E_{\bar{q}}+E'_{\bar{q}})^{2}\lan r^{2}\ran_{20}\ ,\\
\xi'_{G}(1) &=& -\frac{1}{10\sqrt{3}}(E_{\bar{q}}+E'_{\bar{q}})^{2}
\lan r^{2}\ran_{20}
 - \frac{1}{280\sqrt{3}} (E_{\bar{q}}+E'_{\bar{q}})^{4}<r^{4}>_{20}\ .
\label{xigdq}
\eea
\end{itemize}
Note that these expressions include transitions from the ground state
into radially
 excited states. If the two $j=\frac{1}{2}$ states are the same,
$E'_{\bar{q}}=E_{\bar{q}}$ and
$\xi_{C}(1)$ is normalized to one.
Also note that from (\ref{xieq}) and (\ref{xifq}) it follows that
the two $P$  wave form factors
satisfy
\beq
\xi_{E}(\omega) = \frac{\omega+1}{\sqrt{3}}\xi_{F}(\omega)\ ,
\label{pwff}
\eeq
and in particular
\beq
\xi_{E}(1)= \frac{2}{\sqrt{3}}\xi_{F}(1)\ .
\label{pw1}
\eeq

It can be also shown \cite{modelling} that the above formulae
can be
 generalized  to any model involving the
Dirac equation with a spherically symmetric potential.
There, the wave function has the form
\beq
\phi_{j\lambda_{j}}^{(\alpha k)}({\bf x}) =
\left( \begin{array}{c}
f_{\alpha j}^{ k}(r) {\cal Y}_{j\lambda_{j}}^{k}(\Omega)\\
i g_{\alpha j}^{ k}(r) {\cal Y}_{j\lambda_{j}}^{-k}(\Omega)\end{array}
\right)\ ,
\label{wfd}
\eeq
where ${\cal Y}_{j\lambda_{j}}^{k}$ are the usual
spherical spinors,
 $k=l$ ($l=j+\frac{1}{2}$) or $k=-l-1$ $(l=j-\frac{1}{2})$, and
$\alpha$ again denotes all other quantum numbers.
Using (\ref{wfd}) one finds
 that all the expressions
(\ref{xicq})-(\ref{xigdq}) remain unchanged, except for the
expectation value (\ref{jave}) which is
replaced by
\beq
\lan F(r) \ran_{L'L}^{\alpha'\alpha}\rar
\lan F(r) \ran_{j'j}^{\alpha'\alpha} = \int r^{2} dr
[f^{*k'}_{\alpha' j'}(r)f^{k}_{\alpha j}(r) +
g^{*k'}_{\alpha' j'}(r)g^{k}_{\alpha j}(r) ] F(r)\ .
\eeq
Of course, in  models with the Dirac equation
the two $P$  wave doublets are not degenerate any more
($E'_{\bar{q}}\not{\hspace*{-1mm}=\hspace*{-0.01mm}}E_{\bar{q}}$), so that
relations (\ref{pwff}) and (\ref{pw1}) are no longer valid.

\section{Heavy-light models}
\label{models}

Although we have presented here a formalism applicable for a variety
of
transitions $C\rar C,\ldots,G^{*}$, we shall focus
for the rest of the
paper  on the $P$  wave transitions
$B\rar D_{1}l\bar{\nu}_{l}$ and $B\rar D_{2}^{*}l\bar{\nu}_{l}$, and
their $B_{s}$ counterparts. The observed
$D_{1}$ and $D_{2}^{*}$ (or $D_{s1}$ and
$D_{s2}^{*}$) states are expected to be members
of the $j=\frac{3}{2}$ $P$  wave doublet ($F,F^{*}$), since
$j=\frac{1}{2}$ $P$  wave doublet ($E,E^{*}$) will have
$S$  wave decays, and therefore these states should be broad resonances
and correspondingly much harder to observe.
In order to examine the sensitivity of our form factor
predictions to the choice of a specific model and its parameters,
we have performed calculations of $\xi_{F}$ using
four different models in the heavy-quark limit:
 ISGW model, semi-relativistic quark model (SRQM), Dirac equation
with scalar confinement (DESC), and  Salpeter  equation with
vector confinement (SEVC). For the ISGW model we have used
 the original parameters from \cite{isgw}. For the other
three models, we have fixed the string tension $b$
in order to reproduce the expected linear
Regge behavior, and (for a given light quark mass
$m_{u,d}$), varied the other parameters until a good description of
the observed heavy-light mesons was obtained. It turns out
that our results are not sensitive to the exact
value of the light quark mass. By varying
$m_{u,d}$ in the range from $0.300\ GeV$ to $0.350\ GeV$
we have found that  specific choice of $0.300\ GeV$,
that we made for the SRQM, DESC and SEVC, does not
lead to significant uncertainties, as long as
good description
of the observed spin-averaged spectrum of heavy-light states is obtained.
We have also used a single
pseudo-Coulombic (PC) \cite{weniger}
or harmonic oscillator (HO) basis
 wave functions
to obtain form factor predictions. In this case
the unknown energy of the LDF
was estimated from
 the recent experimental data \cite{cleo} for
$B\rar D^{(*)}l\bar{\nu}_{l}$ decays, and also from the
spin-averaged masses
of the known heavy-light states.

\subsection{ISGW model}

Because of its simplicity, the
ISGW model \cite{isgw} is widely used in combination with
HQET for calculations of different form factors. It is a
non-relativistic quark model based on the Schr$\ddot{\rm o}$dinger
equation with the usual Coulomb plus linear potential,
\beq
V(r) = -\frac{4\alpha_{s}}{3r} + c + b r\ .
\eeq
The Hamiltonian of the LDF is then given by
\beq
H_{\bar{q}}= \frac{{\bf p}^{2}}{2\mu} + m_{\bar{q}} + V(r)\ ,
\eeq
with $\mu = \frac{m_{\bar{q}}m_{Q}}{m_{\bar{q}}+m_{Q}}$.

We have used the original parameters from \cite{isgw},
\bea
m_{u,d}&=&0.33 GeV\ ,\nonumber \\
m_{s} &=& 0.55 GeV\ ,\nonumber \\
m_{c}&=&1.82 GeV\ ,\nonumber \\
m_{b} &=& 5.12 GeV\ ,\\
\alpha_{s} &=& 0.50\ ,\nonumber\\
c&=&-0.84 GeV\ ,\nonumber \\
b&=&0.18 GeV^{2}\ ,\nonumber
\label{fit_isgw}
\eea
to obtain  theoretical predictions for the spin-averaged
heavy-light $B$   and $D$  meson states.
The results are shown in Table \ref{tnrqm}.
Taking into account that the non-relativistic quark model
with linear confinement
does not yield linear Regge trajectories, this model provides
a reasonable description of many of the known heavy-light meson states.
Using the LDF wave functions and energies obtained from the
parameters given above, we show in Figures \ref{ff32_isgw} and
\ref{ff32s_isgw} (with full lines)
our predictions (obtained from (\ref{xifq})) for the form factor
$\xi_{F}$ in the
semi-leptonic
decays $B\rar D_{1},D_{2}^{*}$ and corresponding
$B_{s}$ decays. For comparison
 we show with dashed lines the corresponding
AOM \cite{mannel,suzuki} form factors  obtained from
(\ref{xifmannel}).

\subsection{Semi-relativistic quark model (SRQM)}

It was observed in \cite{duncan} from Lattice QCD simulations
that the ground state
wave function describing the LDF in heavy-light mesons
is in remarkably good agreement with the wave function that one gets
from the semi-relativistic quark model.
In this model the Hamiltonian describing the LDF is
\beq
H_{\bar{q}} = \sqrt{{\bf p}^{2}+m_{\bar{q}}^{2}} + V(r)\ ,
\eeq
where
\beq
V(r) = -\frac{4\alpha_{s}}{3r} + b r\ .
\label{potential}
\eeq
The SRQM yields linear Regge trajectories with  slopes of
$\alpha'_{HL}=\frac{1}{4b}$. The  slope of the Regge trajectories
 in the
heavy-light case is expected to
be  exactly twice the slope  in the light-light case \cite{goebel,ovw},
i.e. $\alpha'_{HL} = 2\alpha'_{LL}$.  The observed
Regge slope for the light-light states is $\alpha'_{LL} = 0.88\ GeV^{-2}$
\cite{barger}.
Therefore, in order to obtain the expected Regge behavior, we
fix the string tension $b$ to be\footnote{Although this method
of choosing the effective string tension ensures the correct
Regge behavior, it may  not correspond to the correct static
string tension. If it does not, it indicates the interaction
dynamics is incorrect.}
\bea
b& =& \frac{1}{4 \alpha'_{HL}} \nonumber \\
&=& 0.142\ GeV^{2}\ .
\eea
For a given $m_{u,d}$ we vary the other parameters of the model
to account for all observed heavy-light $B$   and $D$  meson states.
An example of such  fit is given in Table \ref{tsrqm}, with parameters
\bea
m_{u,d}&=&0.300 GeV\ \ \ {\rm (fixed)}\ ,\nonumber \\
m_{s} &=& 0.512 GeV\ ,\nonumber \\
m_{c}&=&1.437 GeV\ ,\nonumber \\
m_{b} &=& 4.774 GeV\ ,
\label{fit_srqm}\\
\alpha_{s} &=& 0.421\ ,\nonumber\\
b&=&0.142 GeV^{2}\ \ \ {\rm (fixed)}\ .\nonumber
\eea
As one can see, the agreement of theoretical and experimental
results is excellent.  We show with dotted lines in Figures
\ref{ff32_mod} and \ref{ff32s_mod} the form factors
for the decays
$B\rar D_{1},D_{2}^{*}$ and
$B_{s}\rar D_{s1},D_{s2}^{*}$, respectively.
{}From these two figures one can see
 that form factors for the two decays
are almost identical. The near equality of the $B$ and $B_{s}$
form factors is a reflection of the similarity of the wave functions
for mesons with or without a strange quark. This in turns
explains the near equality \cite{pdg} of the $m_{D^{*}}-m_{D}
\simeq m_{D^{*}_{s}}-m_{D_{s}}$
(or
$m_{B^{*}}-m_{B}\simeq m_{B^{*}_{s}}-m_{B_{s}}$)
hyperfine differences.

\subsection{Dirac equation with scalar confinement (DESC)}

Scalar confinement is the only type of confinement potential
that has correct sign of the spin-orbit coupling. In the Dirac equation
it also yields linear Regge trajectories, but with  slope
of $\alpha'_{HL}=\frac{1}{2b}$, and one can also
obtain very good description of the spin averaged heavy-light states
\cite{ovw}. In this model the LDF Hamiltonian is given by
\beq
H_{\bar{q}} = H_{0} +  \beta b r
-\frac{4\alpha_{s}}{3r}\ ,\label{deq}
\eeq
where $H_{0}$ is the free particle Dirac Hamiltonian,
\beq
H_{0} = \balpha \cdot {\bf p} + \beta m_{\bar{q}}\ .
\eeq
 Reduction of (\ref{deq})
to the set of radial equations is standard \cite{gross},
and the method of  solution is described in \cite{ovw}.
In order to have the expected Regge behavior, we fix $b$
to
\bea
b& =& \frac{1}{2 \alpha'_{HL}} \nonumber \\
&=& 0.284\ GeV^{2}\ ,
\eea
and, for a given $m_{u,d}$, we vary the other parameters of the model
to account for all observed heavy-light $B$   and $D$  meson states.
In Table \ref{tdesc} we show an example of such fit, with parameters
\bea
m_{u,d}&=&0.300 GeV\ \ \ {\rm (fixed)}\ ,\nonumber \\
m_{s} &=& 0.465 GeV\ ,\nonumber \\
m_{c}&=&1.357 GeV\ ,\nonumber \\
m_{b} &=& 4.693 GeV\ ,
\label{fit_desc}\\
\alpha_{s} &=& 0.462\ ,\nonumber\\
b&=&0.284 GeV^{2}\ \ \ {\rm (fixed)}\ .\nonumber
\eea
The agreement of theoretical and experimental
results is again very good.
The form factors for the decays
$B\rar D_{1},D_{2}^{*}$ and
$B\rar D_{s1},D_{s2}^{*}$,
resulting from the DESC model, with
parameters given above,  are shown with dashed lines in
Figures \ref{ff32_mod} and \ref{ff32s_mod}.
 Note
that form factors obtained from this model are very similar to the
ones obtained from SRQM.

\subsection{Salpeter equation with vector confinement (SEVC)}

The instantaneous version of the Bethe-Salpeter
equation \cite{bethe,gellmann}
(usually referred to as the Salpeter equation \cite{salpeter})
is widely used for the discussion of bound state problems.
It is also equivalent \cite{long}
to the so called ``no-pair'' equation \cite{sucher}, which
was
introduced  in order to avoid the problem of mixing of
positive and negative energy states that occurred in the
Dirac equation for the helium atom. A similar
problem also occurs for a single fermionic particle moving in the
confining Lorentz vector potential. For a very long time \cite{plesset}
it has been known that there are no normalizable solutions
to the Dirac equation in this case.

It has been shown
 analytically  for the heavy-light case \cite{ovw},
and numerically for the case of fermion and antifermion with
arbitrary mass \cite{lagae,ovw2}, that in this type
of model linear scalar confinement does not yield
linear Regge trajectories.
We have therefore used vector confinement, even though
it is well known to give the wrong sign of the spin-orbit coupling.
In terms of the free particle Dirac Hamiltonian
$H_{0}$, potential $V(r)$ from
(\ref{potential}),
 and the positive energy projection operator $\Lambda_{+}$
defined as ($E_{0} = \sqrt{{\bf p}^{2}+m_{\bar{q}}^{2}}$)
\beq
\Lambda_{+} = \frac{E_{0} + H_{0}}
{2E_{0}}\ ,
\eeq
the LDF Hamiltonian for the heavy-light Salpeter equation with
vector confinement is given by
\beq
H_{\bar{q}} = H_{0}
+ \Lambda_{+} V(r)
\Lambda_{+}\ .
\label{npeq}
\eeq
The reduction of (\ref{npeq}) to a pair of coupled radial equations,
as well as the solution method, is described in \cite{ovw}.
In Table \ref{tnpvc} we show an example of theoretical prediction
for the spectrum of spin-averaged heavy-light states. As in the
case of SRQM and DESC, the agreement of theory and experiment
is excellent. The parameters
of the model were
\bea
m_{u,d}&=&0.300 GeV\ \ \ {\rm (fixed)}\ ,\nonumber \\
m_{s} &=& 0.598 GeV\ ,\nonumber \\
m_{c}&=&1.404 GeV\ ,\nonumber \\
m_{b} &=& 4.739 GeV\ ,
\label{fit_npvc}\\
\alpha_{s} &=& 0.534\ ,\nonumber\\
b&=&0.142 GeV^{2}\ \ \ {\rm (fixed)}\ .\nonumber
\eea
Here, $b$ was again fixed to $0.142\ GeV^{2}$ since the model
yields the Regge slope of $\alpha'_{HL}=\frac{1}{4b}$, as in the case
of SRQM.
We again calculate the form factors
resulting from the above parameters. Results  for the decays
$B\rar D_{1},D_{2}^{*}$ and
$B\rar D_{s1},D_{s2}^{*}$, are shown with full lines in
Figures \ref{ff32_mod} and \ref{ff32s_mod}, respectively.
Form factors obtained from this model are
about 10\% larger than the ones obtained
from the SRQM and the DESC.

\subsection{One basis state estimates}
Quite often (as was done in \cite{mannel,wambach}) one
 finds in the literature
 estimates for  form factors
that use a single basis state as a wave function of the LDF.
Usually, it is argued on the basis of the original ISGW model
\cite{isgw}, that it should be the lowest harmonic oscillator
(HO) wave
function with the scale parameter around $0.4\ GeV$.
However, on the basis of  lattice data \cite{duncan}
one might argue that pseudo-Coulombic (PC) basis states
\cite{weniger} are more suitable
for such a purpose. We show in Figure \ref{lattice}
a comparison of the lattice wave function for the heavy-light
system calculated in
\cite{duncan}, with both PC (full line) and
HO $1S$ radial wave function with the scale
parameter  $\beta_{S}=0.40\ GeV$.

Once we choose the LDF wave function, there are still
two unknown parameters ($E_{\bar{q}}$ and
$E'_{\bar{q}}$) in the expression (\ref{xifq}) needed
for the calculation of $\xi_{F}$. In
\cite{one} the energy of the LDF was estimated
by comparing the theoretical prediction for the IW function
$\xi_{C}$ obtained from
(\ref{xicq})   with the recent experimental data \cite{cleo} for
the exclusive semi-leptonic
$B\rar D^{(*)}l\bar{\nu}_{l}$ decay. It is straightforward
to repeat the same analysis using only one basis state, and
obtain the range of acceptable values for $E_{\bar{q}}$ in
$1S$ state ($B,B^{*}$ or $D,D^{*}$ mesons).
The corresponding value for the other $1S$ ($B_{s},B^{*}_{s}$ mesons)
or $1P$ ($D_{1},D^{*}_{2}$ and $D_{s1},D_{s2}^{*}$ mesons)
doublets is then
determined by the difference between the spin averaged
masses of the doublet with unknown
$E_{\bar{q}}$ and the one where
$E_{\bar{q}}$ is known ($B,B^{*}$ or $D,D^{*}$ doublets).
In the following we present the necessary formulae for this
analysis for both PC and HO
wave functions.

\subsubsection{ Pseudo-Coulombic basis states (PC)}

The lowest $1S$ and $1P$ wave functions are \cite{weniger}
\bea
R_{1S}(r) &=& 2\beta_{S}^{3/2}
\exp{(-\beta_{S}r)}\ ,\\
R_{1P}(r) &=& \frac{2}{\sqrt{3}}\beta_{P}^{5/2}r
\exp{(-\beta_{P}r)}\ .
\eea
{}From (\ref{xicq}) we find (with $E_{\bar{q}}=E'_{\bar{q}}$)
\beq
\xi_{C}(\omega) = \frac{2\beta_{S}^{4}(\omega+1)}{((\omega+1)
\beta_{S}^{2}
+(\omega-1) E_{\bar{q}}^{2}))^{2}}\ .
\eeq
Using this expression for $\xi_{C}$ with $\beta_{S} = 0.40\ GeV$
(the corresponding wave function is shown with full line
in Figure \ref{lattice}),
and
performing the analysis as described in \cite{one},
we find that the lowest $\chi^{2}$ of 0.372 per degree of freedom
is obtained for
\beq
E_{\bar{q}}=0.320\ GeV \ ,
\label{pceq1}
\eeq
for the ($B,B^{*}$) and ($D,D^{*}$) doublets.
Adding spin-averaged mass differences given in
Tables \ref{tnrqm}-\ref{tnpvc} we find
\bea
E_{\bar{q}}&=&0.415\ GeV\ ,\\
E_{\bar{q}}&=&0.792\ GeV\ ,\\
E_{\bar{q}}&=&0.905\ GeV\ ,
\label{pceq4}
\eea
for the ($B_{s},B^{*}_{s}$), ($D_{1},D^{*}_{2}$), and
($D_{s1},D^{*}_{s2}$) doublets, respectively.
Using these values for $E_{\bar{q}}$,
and assuming $\beta_{P}=\beta_{S}=0.40\ GeV$ in the expressions valid
for $C\rar F,F^{*}$ transitions,
\bea
\xi_{F}(\omega)
&=& 64\frac{ \beta_{P}^{5/2}\beta_{S}^{3/2}(\beta_{S}+\beta_{P})
(E_{\bar{q}}+E'_{\bar{q}})(\omega +1)}
{((\omega +1)(\beta_{S}+\beta_{P})^{2} + (\omega -1)
(E_{\bar{q}}+E'_{\bar{q}} )^{2})^{3}}\ ,
\label{pcxif}\\
\xi_{F}(1) &=&16\frac{ \beta_{P}^{5/2}\beta_{S}^{3/2}}
{(\beta_{S}+\beta_{P})^{5}}
(E_{\bar{q}}+E'_{\bar{q}} )\ ,\\
\xi'_{F}(1) &=&-16\frac{ \beta_{P}^{5/2}\beta_{S}^{3/2}}
{(\beta_{S}+\beta_{P})^{5}}
(E_{\bar{q}}+E'_{\bar{q}} )\left[1 + \frac{3}{2}
\frac{(E_{\bar{q}}+E'_{\bar{q}} )^2}{(\beta_{S}+\beta_{P})^{2}}
\right]\ ,
\eea
we calculate form factors for $B\rar D_{1},D_{2}^{*}$ and
$B_{s}\rar D_{s1},D_{s2}^{*}$ decays, shown with the full lines
in Figures
\ref{1stb} and \ref{1stbs}, respectively.

\subsubsection{ Harmonic oscillator basis states (HO)}

Here, the lowest $1S$ and $1P$ states,
\bea
R_{1S}(r) &=& \frac{2\beta_{S}^{3/2}}{\pi^{1/4}}
\exp{(-\beta_{S}^{2}r^{2}/2)}\ ,\\
R_{1P}(r) &=& \sqrt{\frac{8}{3}}\frac{\beta_{P}^{5/2}}{\pi^{1/4}}
r \exp{(-\beta_{P}^{2}r^{2}/2)}\ ,
\eea
used in (\ref{xicq}) give (with $E_{\bar{q}}=E'_{\bar{q}}$)
\beq
\xi_{C}(\omega) = \frac{2}{\omega+1}
\exp{\left[- \frac{E_{S}^{2}(\omega-1)}{\beta_{S}^{2}(\omega+1)}
\right]} \ .
\eeq
Again, using this expression for $\xi_{C}$ with $\beta_{S}=0.40\ GeV$,
and performing the analysis from \cite{one}, we find
that
\beq
E_{\bar{q}}=0.444\ GeV\ ,
\label{hoeq1}
\eeq
(for the ($B,B^{*}$) and ($D,D^{*}$) doublets),
 yields the lowest $\chi^{2}$ of
0.357 per degree of freedom
(the corresponding wave function is shown with the dashed line
in Figure \ref{lattice}). This implies
\bea
E_{\bar{q}}&=&0.539\ GeV\ ,\\
E_{\bar{q}}&=&0.916\ GeV\ ,\\
E_{\bar{q}}&=&1.029\ GeV\ ,
\label{hoeq4}
\eea
for the ($B_{s},B^{*}_{s}$), ($D_{1},D^{*}_{2}$), and
($D_{s1},D_{s2}^{*}$) doublets, respectively.
The harmonic oscillator expressions valid
for $C\rar F,F^{*}$ transitions are
\bea
\xi_{F}(\omega)
&=& 8\frac{ \beta_{P}^{5/2}\beta_{S}^{3/2}}
{(\beta_{S}^{2}+\beta_{P}^{2})^{5/2}}
(E_{\bar{q}}+E'_{\bar{q}} )\frac{1}{(\omega +1)^{2}}
\exp{\left[-\frac{
(E_{\bar{q}}+E'_{\bar{q}} )^{2}(\omega-1)}
{2(\beta_{S}^{2}+\beta_{P}^{2})(\omega +1)}\right]}\ ,
\label{hoxif}\\
\xi_{F}(1) &=&2\frac{ \beta_{P}^{5/2}\beta_{S}^{3/2}}
{(\beta_{S}^{2}+\beta_{P}^{2})^{5/2}}
(E_{\bar{q}}+E'_{\bar{q}} )\ ,\\
\xi'_{F}(1) &=&-2\frac{ \beta_{P}^{5/2}\beta_{S}^{3/2}}
{(\beta_{S}^{2}+\beta_{P}^{2})^{5/2}}
(E_{\bar{q}}+E'_{\bar{q}} )\left[1 + \frac{1}{4}
\frac{(E_{\bar{q}}+E'_{\bar{q}} )^2}{(\beta_{S}^{2}+\beta_{P}^{2})}
\right]\ .
\eea
These formulae,
with $\beta_{P}=\beta_{S}=0.40\ GeV$ and $E_{\bar{q}}$
values given above, yield form factors shown with the dashed lines
in Figures
\ref{1stb} and \ref{1stbs}
 for $B\rar D_{1},D_{2}^{*}$ and
$B_{s}\rar D_{s1},D_{s2}^{*}$ decays, respectively.

\section{Results for the decay rates and branching ratios}
\label{results}

For calculation of the decay rates we have chosen $V_{cb}=0.040$
as the reference value. Also, we have used
 \cite{pdg} $\tau_{B}^{ref}=1.50\times 10^{-12}s$
(for $B\rar D_{1},D_{2}^{*}$ decays),
and $\tau_{B_{s}}^{ref}=1.34\times
10^{-12}s$ (for $B_{s}\rar D_{s1},D_{s2}^{*}$ decays).
In order to examine sensitivity of form factors to the
choice of parameters of a specific model,
we have fixed $m_{c}$ between $1.2\ GeV$ and $1.6\ GeV$,
and varied other parameters of SRQM, DESC, and SEVC,
until a good description of
the spin-averaged spectra is obtained. For the one basis state
estimates we have performed the analysis from \cite{one} and
obtained ranges of acceptable $E_{\bar{q}}$ values
($(B,B^{*})$ and $(D,D^{*})$ doublets)
for each
$\beta_{S}$ in the range from $0.3\ GeV$ to $0.5\ GeV$. For
example, for $\beta_{S}=0.3\ GeV$ we found that
$E_{\bar{q}}$ for $(B,B^{*})$ and $(D,D^{*})$ doublets
 ranges from $0.208\ GeV$ to $0.271\ GeV$ in the PC case,
while the corresponding range in the HO case was $0.288\ GeV$
to $0.374\ GeV$. Similarly,
for $\beta_{S}=0.5\ GeV$ the results were
$0.346\ GeV$ to $0.451\ GeV$ in the PC case, and
$0.480\ GeV$
to $0.623\ GeV$ in the HO case. By adding the appropriate
spin-averaged mass differences we obtained values of $E_{\bar{q}}$
for the other doublets.
Assuming $\beta_{P}=\beta_{S}$, and varying $\beta_{S}$
in the range from $0.3\ GeV$ to $0.5\ GeV$ (and $E_{\bar{q}}$
in the range corresponding to a given $\beta_{S}$), we obtained
the acceptable ranges for all decay
rates and branching ratios considered in
this paper.

All results
obtained from different
models and one basis state estimates (for $\xi_{F}(1)$, $\xi'_{F}(1)$,
decay rates and branching ratios),
  are collected in Tables \ref{trb1}-\ref{trbs2}.
As one can
see, the uncertainty introduced by the choice of parameters
within a specific model is the largest for the SRQM (about 30\%),
and the smallest for the estimates which use pseudo-Coulombic basis
states (only about 5\%). For other models the uncertainty
is about (15-25)\%. We also note that the HO wave function estimates
(which are the most commonly encountered in the literature)
yield  branching ratios that are significantly larger
than those obtained from the more realistic models,
for all decays considered
in this paper.

{}From the three models SRQM, DESC and SEVC,
which successfully account for the
known heavy-light masses,
we obtain the following
ranges for the $S$   to $P$  wave decay rates:
\bea
\Gamma (B\rar D_{1}l\bar{\nu}_{l}) &=& (1.16\pm 0.38)\times
\left|\frac{V_{cb}}{0.040}\right|^{2}10^{-15}GeV\ ,\\
\Gamma (B\rar D_{2}^{*}l\bar{\nu}_{l}) &=& (1.96\pm 0.66)\times
\left|\frac{V_{cb}}{0.040}\right|^{2}10^{-15}GeV\ ,\\
\Gamma (B_{s}\rar D_{s1}l\bar{\nu}_{l}) &=& (1.08\pm 0.35)\times
\left|\frac{V_{cb}}{0.040}\right|^{2}10^{-15}GeV\ ,\\
\Gamma (B_{s}\rar D_{s2}^{*}l\bar{\nu}_{l}) &=& (1.74\pm 0.62)\times
\left|\frac{V_{cb}}{0.040}\right|^{2}10^{-15}GeV\ .
\eea
Corresponding branching ratios are (using $V_{cb}=0.040$):
\bea
BR (B\rar D_{1}l\bar{\nu}_{l}) &=& (0.27\pm 0.08)
\frac{\tau_{B}}{1.50ps}\%\ ,
\label{brd1}\\
BR (B\rar D_{2}^{*}l\bar{\nu}_{l}) &=& (0.45\pm 0.14)
\frac{\tau_{B}}{1.50ps}\%\ ,
\label{brd2}\\
BR (B_{s}\rar D_{s1}l\bar{\nu}_{l}) &=&(0.22\pm 0.07)
\frac{\tau_{B_{s}}}{1.34ps}\%\ ,\\
BR (B_{s}\rar D_{s2}^{*}l\bar{\nu}_{l}) &=& (0.36\pm 0.13)
\frac{\tau_{B_{s}}}{1.34ps}\%\ .
\eea

For comparison with an earlier work, we quote results from
\cite{suzuki}, where  $B$  meson decays into charmed
higher resonances were
considered:
\bea
\Gamma (B\rar D_{1}l\bar{\nu}_{l}) &=& 0.36\times
\left|\frac{V_{cb}}{0.040}\right|^{2}10^{-15}GeV\ ,\\
\Gamma (B\rar D_{2}^{*}l\bar{\nu}_{l}) &=& 0.52\times
\left|\frac{V_{cb}}{0.040}\right|^{2}10^{-15}GeV\ ,\\
BR (B\rar D_{1}l\bar{\nu}_{l}) &=& 0.08
\frac{\tau_{B}}{1.50ps}\%\ ,\\
BR (B\rar D_{2}^{*}l\bar{\nu}_{l}) &=& 0.12
\frac{\tau_{B}}{1.50ps}\%\ .
\eea
Let us also mention a few earlier calculations of the $\xi_{E}$ and
$\xi_{F}$ form factors.
In \cite{wambach} these form factors were computed to order
${\cal O}((\omega-1)^{2})$,
\bea
\xi_{E}(\omega) &=& (1.43\pm 0.13) -(1.86\pm 0.28)(\omega-1)
+ {\cal O}((\omega-1)^{2})\ ,\\
\xi_{F}(\omega) &=& (1.14\pm 0.04) -(2.20\pm 0.16)(\omega-1)
+ {\cal O}((\omega-1)^{2})\ ,
\eea
where the quoted
errors are due to a 12\% variation over the scale parameter
of the HO wave function. This should be compared with our
HO estimates for $\xi_{F}(1)$ and $\xi_{F}'(1)$
 given in Tables \ref{trb1} and \ref{trb2} (our errors
are due to 25\% variation of $\beta_{S}=\beta_{P}=0.4$).
Assuming that the $E_{\bar{q}}$ value for the $j=\frac{1}{2}$ $P$
wave doublet is the same as the one for the $j=\frac{3}{2}$ $P$
wave doublet\footnote{This assumption is valid for the NRQM and
the SRQM, which do not distinguish between doublets of the same
orbital
angular momentum but with
different  $j$. However, in the spirit of the one basis state
analysis described above, where experimental mass splitting
between doublets is used for determination of $E_{\bar{q}}$,
$E_{\bar{q}}$'s for the two $P$ wave doublets will be slightly
different.},
our prediction
for $\xi_{E}$ can be obtained using (\ref{pwff}) and (\ref{pw1}).
In particular, using $\xi_{F}(1)=1.23$, we find $\xi_{E}(1)=1.42$.
Note that models based on the Dirac equation can result
in $\xi_{E}$ being larger than $\xi_{F}$.
In \cite{blok}  the QCD sum rule calculation yielded
$\xi_{E}(1)=1.2\pm 0.7$ ($\xi_{F}$ was not determined), while
a Bethe-Salpeter approach of \cite{dai} resulted in
$\xi_{E}(1)=0.73$ and $\xi_{F}(1)=0.76$ \

Finally, in Figures \ref{dgdwf1b} and
\ref{dgdwf2b}
we show differential branching ratios $\frac{dBR}{d\omega}$
for the decays $B\rar D_{1}l\bar{\nu}_{l}$
and $B\rar D_{2}^{*}l\bar{\nu}_{l}$, respectively,
obtained from the four
different models
used in this paper.
Differential branching ratios for the corresponding $B_{s}$ decays
are shown in Figures \ref{dgdwf1bs} and
\ref{dgdwf2bs}. These
calculations assumed the model parameters given in
(\ref{fit_isgw}), (\ref{fit_srqm}), (\ref{fit_desc}) and
(\ref{fit_npvc}).
We also assumed $V_{cb}= 0.040$, $\tau_{B}=1.50
\times 10^{-12}s$ and $\tau_{B_{s}}=1.34
\times 10^{-12}s$. Differential branching ratios
resulting from the one basis state estimates can be
obtained using expressions (\ref{pcxif}) and (\ref{hoxif}), with
parameters given in (\ref{pceq1})-(\ref{pceq4})
and (\ref{hoeq1})-(\ref{hoeq4}).

\section{Conclusion}
\label{con}

In this paper we  have considered semi-leptonic
 $B$   and $B_{s}$  meson decays into the observed charmed
$P$  wave states, in the limit where both $b$   and $c$ quarks
are considered heavy.
We have estimated the unknown form factors in terms
of  overlaps of the wave functions describing the
 final and initial states of the light degrees of freedom.
Unlike in  previous work \cite{suzuki}, our
form factor definitions  are consistent with
the covariant trace formalism of HQET.
As a result of this, we find significantly different results
for decay rates and branching ratios for  processes
$B\rar D_{1}l\bar{\nu}_{l}$ and $B\rar D_{2}^{*}l\bar{\nu}_{l}$.

 In order to examine the sensitivity of our results to
the choice of a specific model, we have performed
all calculations using several different
models. By fixing $m_{c}$ in the range
range from $1.2\ GeV$ to $1.6\ GeV$, and varying the other model
parameters until a good description of the
spin-averaged heavy-light spectrum is obtained,
we  have also examined dependence of our results on the choice
of parameters within a specific model. We have also investigated two
examples of
 one basis state calculations. In those cases the uncertainties
were estimated from the range of acceptable
$E_{\bar{q}}$ values consistent with the experimental
data \cite{cleo} for the decay $B\rar D^{(*)}l\bar{\nu}_{l}$
(as in \cite{one}).
This proceedure leads
to the conclusion that
the choice of  parameters
within the model introduces errors at the level of  (5-30)\%.
Although six models in all are considered we should emphasize that
three (SRQM, DESC, and SEVC), are particularly reliable since they
account for the observed $D$ and $B$ spectroscopies in a
very satisfactory manner.
Between these three models, we find that the predictive accuracy
for the unknown form factors
is about 30\%.

The experimental status for $B\rar D_{J} l \bar{\nu}_{l}$
is still uncertain. At present three experimental
groups have results for these decays but with possible
 additional non-charmed
particle(s) $X$. These results are given in Table \ref{texp}
together with our theoretical predictions which assume
$BR(D_{1}\rar D^{*}\pi) = 67\%$ and
$BR(D_{2}^{*}\rar D^{*}\pi) = 20\%$.
As we can see from Table \ref{texp}
our  predicted branching ratios are consistent
with present measurements. In particular, all upper limits are
satisfied, and where branching ratios have
been determined they are somewhat greater than our predictions
(in which there are no additional particles $X$).

Finally, from Tables \ref{trb1} and \ref{trb2} we see that
the sum
of branching ratios into $D_{1}$ and $D_{2}^{*}$ is about $0.72\%$
(from
the three realistic models).
By counting spin states we estimate the total $P$ wave meson
branching ratio ($E,E^{*},F,F^{*})$ is about $1.08\%$.
Thus the $P$ wave states account for about
one third of the missing semi-leptonic $B$ decays.

\vskip 1cm
\begin{center}
ACKNOWLEDGMENTS
\end{center}
One of us (SV) would like to thank J. F. Beacom for many helpful
conversations.
This work was supported in part by the U.S. Department of Energy
under Contract No. DE-FG02-95ER40896 and in part by the University
of Wisconsin Research Committee with funds granted by the Wisconsin Alumni
Research Foundation.

\newpage

\begin{table}
\normalsize
\begin{center}
TABLES
\end{center}
\caption{ Heavy-light spin averaged states.
Theoretical
results are obtained from the ISGW model \protect\cite{isgw}.
Spin-averaged
masses are calculated in the usual way, by taking $\frac{3}{4}$
($\frac{5}{8}$) of
the triplet and $\frac{1}{4}$ ($\frac{3}{8}$) of the singlet mass for the
$S(P)$  waves). }
\vskip 0.2cm
\begin{tabular}{|lcrccccc|}
\hline
\hline
         State
       & \multicolumn{2}{c}{}
       & Spin-averaged
       & \multicolumn{2}{c}{Q. n.}
       & Theory
       & Error
\\
       & $J^{P}$
       & $^{2S+1}L_{J}$
       & mass (MeV)
       & $j$
       & $k$
       & (MeV)
       & (MeV)
\\
\hline
         \underline{$c\bar{u},\ c\bar{d}$ quarks}
       &
       &
       &
       &
       &
       &
       &
\\
         $\begin{array}{ll}
              		D       (1867) &C \\
   			D^{*}  (2009) &C^{*}\end{array}$
       & $\begin{array}{l}
       			0^{-} \\
      			1^{-} \end{array}$
       &
	 $\left. \begin{array}{l}
			\hspace{+1.1mm}   ^{1}S_{0} \\
			\hspace{+1.1mm}   ^{3}S_{1} \end{array}\right] $
       & $1S\ (1974)$
       & $\frac{1}{2}$
       & $-1$
       & $1931$
       & \hspace*{-5pt} $-43$\hspace*{-5pt}
\\
	 $\begin{array}{ll}
			D_{1}      (2425) &F\\
			D_{2}^{*}  (2459) &F^{*}\end{array}$
       & $\begin{array}{l}
			1^{+} \\
			2^{+} \end{array}$
       & $\left.\begin{array}{r}
			\hspace{+0.5mm}^{1}P_{1}/^{3}P_{1} \\
			\hspace{+0.5mm}^{3}P_{2} \end{array}\right] $
       & $1P\ (2446)$
       & $\frac{3}{2}$
       & $-2$
       & $2447$
       & \hspace*{-5pt} $1$ \hspace*{-5pt}
\\
         \underline{$c\bar{s}$ quarks}
       &
       &
       &
       &
       &
       &
       &
\\
	 $\begin{array}{ll}
   			D_{s}  (1969) &C\\
   			D^{*}_{s}  (2112)&C^{*} \end{array}$
       & $\begin{array}{l}
      			0^{-} \\
      			1^{-} \end{array}$
       & $\left. \begin{array}{l}
			\hspace{+1.1mm}    ^{1}S_{0} \\
			\hspace{+1.1mm}    ^{3}S_{1} \end{array}\right] $
       & $1S\ (2076)$
       & $\frac{1}{2}$
       & $-1$
       & $1982$
       & \hspace*{-5pt} $-94$ \hspace*{-5pt}
\\
	 $\begin{array}{ll}
			D_{s1}  (2535)&\hspace*{-1.5mm} F \\
		        D_{s2}^{*}  (2573)&\hspace*{-1.5mm} F^{*} \end{array}$
       & $\begin{array}{l}
			1^{+} \\
			2^{+} \end{array}$
       & $\left.\begin{array}{r}
			\hspace{+0.5mm}^{1}P_{1}/^{3}P_{1} \\
		  	\hspace{+0.5mm}^{3}P_{2} \end{array}\right] $
       & $1P\ (2559)$
       & $\frac{3}{2}$
       & $-2$
       & $2473$
       & \hspace*{-5pt} $-86$ \hspace*{-5pt}
\\
         \underline{$b\bar{u},\ b\bar{d}$ quarks}
       &
       &
       &
       &
       &
       &
       &
\\
         $\begin{array}{ll}
   			B      (5279) &C\\
   			B^{*}   (5325) &C^{*}\end{array}$
       & $\begin{array}{l}
     	 		0^{-} \\
      			1^{-} \end{array}$
       & $\left. \begin{array}{l}
			\hspace{+1.1mm}   ^{1}S_{0} \\
 			\hspace{+1.1mm}   ^{3}S_{1} \end{array}\right] $
       & $1S\ (5314)$
       & $\frac{1}{2}$
       & $-1$
       & $5188$
       & \hspace*{-5pt} $-126$ \hspace*{-5pt}
\\
         \underline{$b\bar{s}$ quarks}
       &
       &
       &
       &
       &
       &
       &
\\
         $\begin{array}{ll}
			B_{s}     (5374) &C\\
			B_{s}^{*} (5421) &C^{*}\end{array}$
       & $\begin{array}{l}
			0^{-} \\
			1^{-} \end{array} $
       & $\left.\begin{array}{l}
			\hspace{+1.0mm} ^{1}S_{0} \\
			\hspace{+1.0mm} ^{3}S_{1} \end{array}\right] $
       & $1S\ (5409)$
       & $\frac{1}{2}$
       & $-1$
       & $5216$
       & \hspace*{-5pt} $-193$ \hspace*{-5pt}
\\
\hline
\hline
\end{tabular}
\\
\label{tnrqm}
\end{table}

\begin{table}
\caption{ Heavy-light spin averaged states.
Theoretical
results are obtained from the SRQM.
Parameters of the model are given
in (\protect\ref{fit_srqm}).
Spin-averaged
masses are calculated in the usual way, by taking $\frac{3}{4}$
($\frac{5}{8}$) of
the triplet and $\frac{1}{4}$ ($\frac{3}{8}$) of the singlet mass for the
$S(P)$  waves). }
\vskip 0.2cm
\begin{tabular}{|lcrccccc|}
\hline
\hline
         State
       & \multicolumn{2}{c}{Spectroscopic label\hspace{+2mm}}
       & Spin-averaged
       & \multicolumn{2}{c}{Q. n.}
       & Theory
       & Error
\\
       & $J^{P}$
       & $^{2S+1}L_{J}$
       & mass (MeV)
       & $j$
       & $k$
       & (MeV)
       & (MeV)
\\
\hline
         \underline{$c\bar{u},\ c\bar{d}$ quarks}
       &
       &
       &
       &
       &
       &
       &
\\
         $\begin{array}{ll}
              		D       (1867) &C\\
   			D^{*}   (2009) &C^{*}\end{array}$
       & $\begin{array}{l}
       			0^{-} \\
      			1^{-} \end{array}$
       &
	 $\left. \begin{array}{l}
			\hspace{+1.1mm}   ^{1}S_{0} \\
			\hspace{+1.1mm}   ^{3}S_{1} \end{array}\right] $
       & $1S\ (1974)$
       & $\frac{1}{2}$
       & $-1$
       & $1974$
       & $0$
\\
	 $\begin{array}{ll}
			D_{1}      (2425) &F \\
			D_{2}^{*}  (2459) &F^{*}\end{array}$
       & $\begin{array}{l}
			1^{+} \\
			2^{+} \end{array}$
       & $\left.\begin{array}{r}
			\hspace{+0.5mm}^{1}P_{1}/^{3}P_{1} \\
			\hspace{+0.5mm}^{3}P_{2} \end{array}\right] $
       & $1P\ (2446)$
       & $\frac{3}{2}$
       & $-2$
       & $2449$
       & $3$
\\
         \underline{$c\bar{s}$ quarks}
       &
       &
       &
       &
       &
       &
       &
\\
	 $\begin{array}{ll}
   			D_{s}  (1969) &C\\
   			D^{*}_{s}  (2112)&C^{*} \end{array}$
       & $\begin{array}{l}
      			0^{-} \\
      			1^{-} \end{array}$
       & $\left. \begin{array}{l}
			\hspace{+1.1mm}    ^{1}S_{0} \\
			\hspace{+1.1mm}    ^{3}S_{1} \end{array}\right] $
       & $1S\ (2076)$
       & $\frac{1}{2}$
       & $-1$
       & $2075$
       & $-1$
\\
	 $\begin{array}{ll}
			D_{s1}  (2535) &\hspace*{-1.5mm}F\\
		        D_{s2}^{*}  (2573) &\hspace*{-1.5mm}F^{*}\end{array}$
       & $\begin{array}{l}
			1^{+} \\
			2^{+} \end{array}$
       & $\left.\begin{array}{r}
			\hspace{+0.5mm}^{1}P_{1}/^{3}P_{1} \\
		  	\hspace{+0.5mm}^{3}P_{2} \end{array}\right] $
       & $1P\ (2559)$
       & $\frac{3}{2}$
       & $-2$
       & $2557$
       & $-2$
\\
         \underline{$b\bar{u},\ b\bar{d}$ quarks}
       &
       &
       &
       &
       &
       &
       &
\\
         $\begin{array}{ll}
   			B      (5279) &C \\
   			B^{*}   (5325) &C^{*}\end{array}$
       & $\begin{array}{l}
     	 		0^{-} \\
      			1^{-} \end{array}$
       & $\left. \begin{array}{l}
			\hspace{+1.1mm}   ^{1}S_{0} \\
 			\hspace{+1.1mm}   ^{3}S_{1} \end{array}\right] $
       & $1S\ (5314)$
       & $\frac{1}{2}$
       & $-1$
       & $5311$
       & $-3$
\\
         \underline{$b\bar{s}$ quarks}
       &
       &
       &
       &
       &
       &
       &
\\
         $\begin{array}{ll}
			B_{s}      (5374)&C \\
			B_{s}^{*}  (5421) &C^{*}\end{array}$
       & $\begin{array}{l}
			0^{-} \\
			1^{-} \end{array} $
       & $\left.\begin{array}{l}
			\hspace{+1.0mm} ^{1}S_{0} \\
			\hspace{+1.0mm} ^{3}S_{1} \end{array}\right] $
       & $1S\ (5409)$
       & $\frac{1}{2}$
       & $-1$
       & $5412$
       & $3$
\\
\hline
\hline
\end{tabular}
\\
\label{tsrqm}
\end{table}

\begin{table}
\caption{ Heavy-light spin averaged states.
Theoretical
results are obtained from the Dirac equation with scalar
confinement. Parameters of the model are given
in (\protect\ref{fit_desc}).
Spin-averaged
masses are calculated in the usual way, by taking $\frac{3}{4}$
($\frac{5}{8}$) of
the triplet and $\frac{1}{4}$ ($\frac{3}{8}$) of the singlet mass for the
$S(P)$  waves). }
\vskip 0.2cm
\begin{tabular}{|lcrccccc|}
\hline
\hline
         State
       & \multicolumn{2}{c}{}
       & Spin-averaged
       & \multicolumn{2}{c}{Q. n.}
       & Theory
       & Error
\\
       & $J^{P}$
       & $^{2S+1}L_{J}$
       & mass (MeV)
       & $j$
       & $k$
       & (MeV)
       & (MeV)
\\
\hline
         \underline{$c\bar{u},\ c\bar{d}$ quarks}
       &
       &
       &
       &
       &
       &
       &
\\
         $\begin{array}{ll}
              		D       (1867) &C\\
   			D^{*}   (2009) &C^{*}\end{array}$
       & $\begin{array}{l}
       			0^{-} \\
      			1^{-} \end{array}$
       &
	 $\left. \begin{array}{l}
			\hspace{+1.1mm}   ^{1}S_{0} \\
			\hspace{+1.1mm}   ^{3}S_{1} \end{array}\right] $
       & $1S\ (1974)$
       & $\frac{1}{2}$
       & $-1$
       & $1977$
       & $3$
\\
	 $\begin{array}{ll}
			D_{1}     (2425) &F\\
			D_{2}^{*}  (2459) &F^{*}\end{array}$
       & $\begin{array}{l}
			1^{+} \\
			2^{+} \end{array}$
       & $\left.\begin{array}{r}
			\hspace{+0.5mm}^{1}P_{1}/^{3}P_{1} \\
			\hspace{+0.5mm}^{3}P_{2} \end{array}\right] $
       & $1P\ (2446)$
       & $\frac{3}{2}$
       & $-2$
       & $2444$
       & $-2$
\\
         \underline{$c\bar{s}$ quarks}
       &
       &
       &
       &
       &
       &
       &
\\
	 $\begin{array}{ll}
   			D_{s}  (1969) &C \\
   			D^{*}_{s}  (2112) &C^{*}\end{array}$
       & $\begin{array}{l}
      			0^{-} \\
      			1^{-} \end{array}$
       & $\left. \begin{array}{l}
			\hspace{+1.1mm}    ^{1}S_{0} \\
			\hspace{+1.1mm}    ^{3}S_{1} \end{array}\right] $
       & $1S\ (2076)$
       & $\frac{1}{2}$
       & $-1$
       & $2074$
       & $-2$
\\
	 $\begin{array}{ll}
			D_{s1}  (2535) &\hspace*{-1.5mm}F \\
		        D_{s2}^{*}  (2573) &\hspace*{-1.5mm}F^{*}\end{array}$
       & $\begin{array}{l}
			1^{+} \\
			2^{+} \end{array}$
       & $\left.\begin{array}{r}
			\hspace{+0.5mm}^{1}P_{1}/^{3}P_{1} \\
		  	\hspace{+0.5mm}^{3}P_{2} \end{array}\right] $
       & $1P\ (2559)$
       & $\frac{3}{2}$
       & $-2$
       & $2560$
       & $1$
\\
         \underline{$b\bar{u},\ b\bar{d}$ quarks}
       &
       &
       &
       &
       &
       &
       &
\\
         $\begin{array}{ll}
   			B      (5279) &C \\
   			B^{*}   (5325) &C^{*}\end{array}$
       & $\begin{array}{l}
     	 		0^{-} \\
      			1^{-} \end{array}$
       & $\left. \begin{array}{l}
			\hspace{+1.1mm}   ^{1}S_{0} \\
 			\hspace{+1.1mm}   ^{3}S_{1} \end{array}\right] $
       & $1S\ (5314)$
       & $\frac{1}{2}$
       & $-1$
       & $5313$
       & $-1$
\\
         \underline{$b\bar{s}$ quarks}
       &
       &
       &
       &
       &
       &
       &
\\
         $\begin{array}{ll}
			B_{s}      (5374) &C\\
			B_{s}^{*}  (5421) &C^{*}\end{array}$
       & $\begin{array}{l}
			0^{-} \\
			1^{-} \end{array} $
       & $\left.\begin{array}{l}
			\hspace{+1.0mm} ^{1}S_{0} \\
			\hspace{+1.0mm} ^{3}S_{1} \end{array}\right] $
       & $1S\ (5409)$
       & $\frac{1}{2}$
       & $-1$
       & $5410$
       & $1$
\\
\hline
\hline
\end{tabular}
\\
\label{tdesc}
\end{table}

\begin{table}
\caption{ Heavy-light spin averaged states.
Theoretical
results are obtained from the  Salpeter equation
with vector
confinement (in the heavy-light limit). Parameters of the model are given
in (\protect\ref{fit_npvc}).
Spin-averaged
masses are calculated in the usual way, by taking $\frac{3}{4}$
($\frac{5}{8}$) of
the triplet and $\frac{1}{4}$ ($\frac{3}{8}$) of the singlet mass for the
$S(P)$  waves). }
\vskip 0.2cm
\begin{tabular}{|lcrccccc|}
\hline
\hline
         State
       & \multicolumn{2}{c}{}
       & Spin-averaged
       & \multicolumn{2}{c}{Q. n.}
       & Theory
       & Error
\\
       & $J^{P}$
       & $^{2S+1}L_{J}$
       & mass (MeV)
       & $j$
       & $k$
       & (MeV)
       & (MeV)
\\
\hline
         \underline{$c\bar{u},\ c\bar{d}$ quarks}
       &
       &
       &
       &
       &
       &
       &
\\
         $\begin{array}{ll}
              		D     (1867) & C \\
   			D^{*} (2009) & C^{*}\end{array}$
       & $\begin{array}{l}
       			0^{-} \\
      			1^{-} \end{array}$
       &
	 $\left. \begin{array}{l}
			\hspace{+1.1mm}   ^{1}S_{0} \\
			\hspace{+1.1mm}   ^{3}S_{1} \end{array}\right] $
       & $1S\ (1974)$
       & $\frac{1}{2}$
       & $-1$
       & $1980$
       & $6$
\\
	 $\begin{array}{ll}
			D_{1}      (2425) & F\\
			D_{2}^{*}  (2459) & F^{*}\end{array}$
       & $\begin{array}{l}
			1^{+} \\
			2^{+} \end{array}$
       & $\left.\begin{array}{r}
			\hspace{+0.5mm}^{1}P_{1}/^{3}P_{1} \\
			\hspace{+0.5mm}^{3}P_{2} \end{array}\right] $
       & $1P\ (2446)$
       & $\frac{3}{2}$
       & $-2$
       & $2439$
       & $-7$
\\
         \underline{$c\bar{s}$ quarks}
       &
       &
       &
       &
       &
       &
       &
\\
	 $\begin{array}{ll}
   			D_{s}  (1969) &C \\
   			D^{*}_{s}  (2112)&C^{*} \end{array}$
       & $\begin{array}{l}
      			0^{-} \\
      			1^{-} \end{array}$
       & $\left. \begin{array}{l}
			\hspace{+1.1mm}    ^{1}S_{0} \\
			\hspace{+1.1mm}    ^{3}S_{1} \end{array}\right] $
       & $1S\ (2076)$
       & $\frac{1}{2}$
       & $-1$
       & $2072$
       & $-4$
\\
	 $\begin{array}{ll}
			D_{s1}  (2535) &\hspace*{-1.5mm}F \\
		        D_{s2}^{*}  (2573) &\hspace*{-1.5mm}F^{*}\end{array}$
       & $\begin{array}{l}
			1^{+} \\
			2^{+} \end{array}$
       & $\left.\begin{array}{r}
			\hspace{+0.5mm}^{1}P_{1}/^{3}P_{1} \\
		  	\hspace{+0.5mm}^{3}P_{2} \end{array}\right] $
       & $1P\ (2559)$
       & $\frac{3}{2}$
       & $-2$
       & $2564$
       & $5$
\\
         \underline{$b\bar{u},\ b\bar{d}$ quarks}
       &
       &
       &
       &
       &
       &
       &
\\
         $\begin{array}{ll}
   			B      (5279) &C \\
   			B^{*}   (5325) &C^{*}\end{array}$
       & $\begin{array}{l}
     	 		0^{-} \\
      			1^{-} \end{array}$
       & $\left. \begin{array}{l}
			\hspace{+1.1mm}   ^{1}S_{0} \\
 			\hspace{+1.1mm}   ^{3}S_{1} \end{array}\right] $
       & $1S\ (5314)$
       & $\frac{1}{2}$
       & $-1$
       & $5316$
       & $2$
\\
         \underline{$b\bar{s}$ quarks}
       &
       &
       &
       &
       &
       &
       &
\\
         $\begin{array}{ll}
			B_{s}      (5374) &C \\
			B_{s}^{*}  (5421) &C^{*}\end{array}$
       & $\begin{array}{l}
			0^{-} \\
			1^{-} \end{array} $
       & $\left.\begin{array}{l}
			\hspace{+1.0mm} ^{1}S_{0} \\
			\hspace{+1.0mm} ^{3}S_{1} \end{array}\right] $
       & $1S\ (5409)$
       & $\frac{1}{2}$
       & $-1$
       & $5407$
       & $-2$
\\
\hline
\hline
\end{tabular}
\\
\label{tnpvc}
\end{table}

\begin{table}
\caption{Results for the decay
$B\rar D_{1}l\bar{\nu}_{l}$ obtained from four different models
and two one basis state estimates.
 Errors for the SRQM, DESC, and SEVC, are due
to variation of $m_{c}$ in the range from $1.2\ GeV$ to
$1.6\ GeV$. Errors for the PC and HO estimates
are due to variation of $\beta_{S}=\beta_{P}$ in
the range from $0.3\ GeV$ to $0.5\ GeV$.
For the reference values of the
$B$  meson lifetime we take \protect\cite{pdg}
$\tau_{B}^{ref}=1.50\times 10^{-12}s$, and
 for the
reference value of $V_{cb}$ we take 0.040.}
\label{trb1}
\vskip 0.2cm
\begin{tabular}{|lcccc|}
\hline
\hline
Model  & $\xi_{F}(1) $ &  $\xi'_{F}(1) $ &
$\Gamma\ [\left|\frac{V_{cb}}{0.040}\right|^{2}10^{-15}GeV]$ &
$BR\ [\frac{ \tau_{B}^{ } } { \tau_{B}^{ref} }\%] $ \\
\hline
ISGW
& 0.60 & -0.89 & 0.73 & $0.17$ \\
SRQM
& $0.84\pm 0.19$  & $-2.04\pm 0.77$ & $1.01\pm 0.34$
& $0.23\pm 0.08$ \\
DESC
& $0.79\pm 0.14$
& $-1.45\pm 0.44$ & $1.07\pm 0.24$ & $0.25\pm 0.06$ \\
SEVC
& $1.18\pm 0.19$ & $-4.04\pm 1.31$ &
$1.41\pm 0.22 $& $0.32\pm 0.05$ \\
PC
& $1.42\pm 0.26$ & $-6.23\pm 2.70$ &
$1.61\pm 0.06 $& $0.37\pm 0.02$ \\
HO
& $1.23\pm 0.21$ & $-3.23\pm 1.18$ &
$1.94\pm 0.28 $& $0.44\pm 0.06$ \\
\hline
\hline
\end{tabular}
\end{table}

\begin{table}
\caption{Results for the decay
$B\rar D_{2}^{*}l\bar{\nu}_{l}$ obtained from four different models
and two one basis state estimates.
 Errors for the SRQM, DESC, and SEVC, are due
to variation of $m_{c}$ in the range from $1.2\ GeV$ to
$1.6\ GeV$. Errors for the PC and HO estimates
are due to variation of $\beta_{S}=\beta_{P}$ in
the range from $0.3\ GeV$ to $0.5\ GeV$.
We take \protect\cite{pdg}
$\tau_{B}^{ref}=1.50\times 10^{-12}s$ and $V_{cb}=0.040$.}
\label{trb2}
\vskip 0.2cm
\begin{tabular}{|lcccc|}
\hline
\hline
Model  & $\xi_{F}(1) $ &  $\xi'_{F}(1) $ &
$\Gamma\ [\left|\frac{V_{cb}}{0.040}\right|^{2}10^{-15}GeV]$ &
$BR\ [\frac{ \tau_{B}^{ } } { \tau_{B}^{ref} }\%] $ \\
\hline
ISGW
& 0.60 & -0.89 & 1.14 & $0.26$ \\
SRQM
& $0.84\pm 0.19$  & $-2.04\pm 0.77$ & $1.69\pm 0.56$
& $0.38\pm 0.13$ \\
DESC
& $0.79\pm 0.14$
& $-1.45\pm 0.44$ & $1.77\pm 0.45$ & $0.41\pm 0.11$ \\
SEVC
& $1.18\pm 0.19$ & $-4.04\pm 1.31$ &
$2.41\pm 0.41 $& $0.55\pm 0.10$ \\
PC
& $1.42\pm 0.26$ & $-6.23\pm 2.70$ &
$2.80\pm 0.18 $& $0.64\pm 0.04$ \\
HO
& $1.23\pm 0.21$ & $-3.23\pm 1.18$ &
$3.23\pm 0.55 $& $0.74\pm 0.13$ \\
\hline
\hline
\end{tabular}
\end{table}

\begin{table}
\caption{Results for the decay
$B_{s}\rar D_{s1}l\bar{\nu}_{l}$ obtained from four different models
and two one basis state estimates.
 Errors for the SRQM, DESC, and SEVC, are due
to variation of $m_{c}$ in the range from $1.2\ GeV$ to
$1.6\ GeV$. Errors for the PC and HO estimates
are due to variation of $\beta_{S}=\beta_{P}$ in
the range from $0.3\ GeV$ to $0.5\ GeV$.
We take \protect\cite{pdg}
$\tau_{B_{s}}^{ref}=1.34\times 10^{-12}s$ and
$V_{cb}=0.040$.}
\label{trbs1}
\vskip 0.2cm
\begin{tabular}{|lcccc|}
\hline
\hline
Model  & $\xi_{F}(1) $ &  $\xi'_{F}(1) $ &
$\Gamma\ [\left|\frac{V_{cb}}{0.040}\right|^{2}10^{-15}GeV]$ &
$BR\ [\frac{ \tau_{B_{s}}^{ } } { \tau_{B_{s}}^{ref} }\%] $ \\
\hline
ISGW
& 0.51 & -0.71 & 0.53 & $0.11$ \\
SRQM
& $0.84\pm 0.21$  & $-2.09\pm 0.81$ & $0.98\pm 0.34$
& $0.20\pm 0.07$ \\
DESC
& $0.82\pm 0.14$
& $-1.59\pm 0.47$ & $1.10\pm 0.27$ & $0.23\pm 0.06$ \\
SEVC
& $0.97\pm 0.22$ & $-2.75\pm 1.04$ &
$1.15\pm 0.34 $& $0.24\pm 0.07$ \\
PC
& $1.70\pm 0.34$ & $-9.98\pm 4.73$ &
$1.57\pm 0.11 $& $0.32\pm 0.02$ \\
HO
& $1.42\pm 0.26$ & $-4.59\pm 1.87$ &
$2.08\pm 0.21 $& $0.43\pm 0.05$ \\
\hline
\hline
\end{tabular}
\end{table}

\begin{table}
\caption{Results for the decay
$B\rar D_{s2}^{*}l\bar{\nu}_{l}$ obtained from four different models
and two one basis state estimates.
 Errors for the SRQM, DESC, and SEVC, are due
to variation of $m_{c}$ in the range from $1.2\ GeV$ to
$1.6\ GeV$. Errors for the PC and HO estimates
are due to variation of $\beta_{S}=\beta_{P}$ in
the range from $0.3\ GeV$ to $0.5\ GeV$.
For the reference values of the
We take \protect\cite{pdg}
$\tau_{B_{s}}^{ref}=1.34\times 10^{-12}s$ and $V_{cb}=0.040$.}
\label{trbs2}
\vskip 0.2cm
\begin{tabular}{|lcccc|}
\hline
\hline
Model  & $\xi_{F}(1) $ &  $\xi'_{F}(1) $ &
$\Gamma\ [\left|\frac{V_{cb}}{0.040}\right|^{2}10^{-15}GeV]$ &
$BR\ [\frac{ \tau_{B_{s}}^{ } } { \tau_{B_{s}}^{ref} }\%] $ \\
\hline
ISGW
& 0.51 & -0.71 & 0.81 & $0.17$ \\
SRQM
& $0.84\pm 0.21$  & $-2.09\pm 0.81$ & $1.58\pm 0.57$
& $0.32\pm 0.12$ \\
DESC
& $0.82\pm 0.14$
& $-1.59\pm 0.47$ & $1.75\pm 0.45$ & $0.36\pm 0.09$ \\
SEVC
& $0.97\pm 0.22$ & $-2.75\pm 1.04$ &
$1.90\pm 0.60 $& $0.39\pm 0.12$ \\
PC
& $1.70\pm 0.34$ & $-9.98\pm 4.73$ &
$2.90\pm 0.08 $& $0.59\pm 0.02$ \\
HO
& $1.42\pm 0.26$ & $-4.59\pm 1.87$ &
$3.52\pm 0.48 $& $0.72\pm 0.10$ \\
\hline
\hline
\end{tabular}
\end{table}

\begin{table}
\caption{
A summary of experimental results for $B\rar D_{J}Xe^{-}\bar{\nu}$,
where $X$ is a possible non-charmed hadron. Our theoretical predictions
are obtained from (\protect\ref{brd1}) and (\protect\ref{brd2})
for the pure semi-leptonic decays
$B\rar D_{J}e^{-}\bar{\nu}$, assuming
$BR(D_{1}^{0}\rar D^{*+}\pi^{-}) = 67\%$ and
$BR(D_{2}^{*0}\rar D^{*+}\pi^{-}) = 20\%$.}
\label{texp}
\vskip 0.2cm
\begin{tabular}{|lcccc|}
\hline
\hline
Decay Mode  & CLEO \protect\cite{alex} &
 ALEPH  \protect\cite{aleph} & OPAL \protect\cite{opal}
& This work \\
	    & [\%]  & [\%]
& [\%]  & [\%] \\
\hline
$BR(B\rar D_{1}^{0}Xe^{-}\bar{\nu})$
& $<0.67$ & $0.51\pm 0.17$ & $1.36\pm 0.46 $ &
$0.18\pm 0.05 $ \\
$\times BR(D_{1}^{0}\rar D^{*+}\pi^{-}) $ &
(90\% c.l.) & \multicolumn{3}{c|}{} \\
$BR(B\rar D_{2}^{*0}Xe^{-}\bar{\nu})$
& $<0.79$ & $<0.20$ & $0.18\pm 0.08 $ &
$0.09\pm 0.03 $ \\
$\times BR(D_{2}^{*0}\rar D^{*+}\pi^{-}) $ &
(90\% c.l.) & (95\% c.l.) &  \multicolumn{2}{c|}{} \\
\hline
\hline
\end{tabular}
\vspace*{+8cm}
\end{table}

\clearpage

\newpage

\begin{figure}
\begin{center}
FIGURES
\end{center}
\end{figure}

\begin{figure}
\caption{$\xi_{F}$ for the semi-leptonic
decays $B\rar D_{1},D_{2}^{*}$, obtained from the ISGW model \protect
\cite{isgw}. The full line shows our prediction (VO), obtained from
(\protect\ref{xifq}), while the dashed line is AOM prediction
obtained from (\protect\ref{xifmannel}), which is used in
\protect\cite{mannel,suzuki}.}
\label{ff32_isgw}
\end{figure}

\begin{figure}
\caption{$\xi_{F}$ for the semi-leptonic
decays $B_{s}\rar D_{s1},D_{s2}^{*}$,
obtained from the ISGW model \protect
\cite{isgw}. The full line  shows our prediction (VO),
obtained from
(\protect\ref{xifq}), while the dashed line is AOM prediction
obtained from (\protect\ref{xifmannel}), which is used in
\protect\cite{mannel,suzuki}.}
\label{ff32s_isgw}
\end{figure}

\begin{figure}
\caption{$\xi_{F}$ for the semi-leptonic
decays $B\rar D_{1},D_{2}^{*}$, obtained from the SRQM (dotted line),
DESC (dashed line), and
SEVC (full line). Model parameters are given in (\protect\ref{fit_srqm}),
(\protect\ref{fit_desc}) and (\protect\ref{fit_npvc}), respectively.}
\label{ff32_mod}
\end{figure}

\begin{figure}
\caption{$\xi_{F}$ for the semi-leptonic
decays $B_{s}\rar D_{s1},D_{s2}^{*}$,
obtained from the SRQM (dotted line),
DESC (dashed line), and
SEVC (full line). Model parameters are given in (\protect\ref{fit_srqm}),
(\protect\ref{fit_desc}) and (\protect\ref{fit_npvc}), respectively.}
\label{ff32s_mod}
\end{figure}

\begin{figure}
\caption{Comparison of the lattice data
with the $1S$ pseudo-Coulombic
(full line) and harmonic oscillator (dashed line)
wave function. For both wave functions we used
$\beta_{S}=0.40\ GeV$. }
\label{lattice}
\end{figure}

\begin{figure}
\caption{$\xi_{F}$ for the semi-leptonic
decays $B\rar D_{1},D_{2}^{*}$, obtained from one
PC basis state (full line), and from one HO state
(dashed line). In both cases we used $\beta_{S}=\beta_{P}=0.40\ GeV$.
In the PC case we used $E_{\bar{q}}=0.320\ GeV$
($(B,B^{*})$ doublet) and
$E_{\bar{q}}=0.792\ GeV$ ($(D_{1},D^{*}_{2})$ doublet), and
in the HO case
$E_{\bar{q}}=0.444\ GeV$
($(B,B^{*})$ doublet) and
$E_{\bar{q}}=0.916\ GeV$ ($(D_{1},D^{*}_{2})$ doublet).}
\label{1stb}
\end{figure}

\begin{figure}
\caption{$\xi_{F}$ for the semi-leptonic
decays $B_{s}\rar D_{s1},D_{s2}^{*}$, obtained from  one
PC basis state (full line), and from one HO state
(dashed line). In both cases we used $\beta_{S}=\beta_{P}=0.40\ GeV$.
In the PC case we used $E_{\bar{q}}=0.415\ GeV$
($(B_{s},B^{*}_{s})$ doublet) and
$E_{\bar{q}}=0.905\ GeV$ ($(D_{s1},D_{s2}^{*})$ doublet), and
in the HO case
$E_{\bar{q}}=0.539\ GeV$
($(B_{s},B^{*}_{s})$ doublet) and
$E_{\bar{q}}=1.029\ GeV$ ($(D_{s1},D_{s2}^{*})$ doublet).}
\label{1stbs}
\end{figure}

\begin{figure}
\caption{Differential branching ratio  $\frac{d BR}{d\omega}$
for the process $B\rar D_{1}\l\bar{\nu}_{l}$ ($C\rar F$)
obtained from the four different models, with parameters
given in (\protect\ref{fit_isgw}) (ISGW), (\protect\ref{fit_srqm})
(SRQM), (\protect\ref{fit_desc}) (DESC), and (\protect\ref{fit_npvc})
(SEVC). For this calculation
we used $V_{cb} = 0.040$ and $\tau_{B}= 1.50\times
10^{-12}s$. The kinematic limit for this decay
is $\omega_{max}=1.318$.}
\label{dgdwf1b}
\end{figure}

\begin{figure}
\caption{Differential branching ratio  $\frac{d BR}{d\omega}$
for the process $B\rar D_{2}^{*}\l\bar{\nu}_{l}$ ($C\rar F^{*}$)
obtained from the four different models, with parameters
given in (\protect\ref{fit_isgw}) (ISGW), (\protect\ref{fit_srqm})
(SRQM), (\protect\ref{fit_desc}) (DESC), and (\protect\ref{fit_npvc})
(SEVC). For this calculation
we used $V_{cb} = 0.040$ and $\tau_{B}= 1.50\times
10^{-12}s$. The kinematic limit for this decay
is $\omega_{max}=1.306$.}
\label{dgdwf2b}
\end{figure}

\begin{figure}
\caption{Differential branching ratio  $\frac{d BR}{d\omega}$
for the process $B_{s}\rar D_{s1}\l\bar{\nu}_{l}$ ($C\rar F$)
obtained from the four different models, with parameters
given in (\protect\ref{fit_isgw}) (ISGW), (\protect\ref{fit_srqm})
(SRQM), (\protect\ref{fit_desc}) (DESC), and (\protect\ref{fit_npvc})
(SEVC). For this calculation
we used $V_{cb} = 0.040$ and $\tau_{B_{s}}= 1.34\times
10^{-12}s$. The kinematic limit for this decay
is $\omega_{max}=1.296$.}
\label{dgdwf1bs}
\end{figure}

\begin{figure}
\caption{Differential branching ratio  $\frac{d BR}{d\omega}$
for the process $B_{s}\rar D_{s2}^{*}\l\bar{\nu}_{l}$ ($C\rar F^{*}$)
obtained from the four different models, with parameters
given in (\protect\ref{fit_isgw}) (ISGW), (\protect\ref{fit_srqm})
(SRQM), (\protect\ref{fit_desc}) (DESC), and (\protect\ref{fit_npvc})
(SEVC). For this calculation
we used $V_{cb} = 0.040$ and $\tau_{B_{s}}= 1.34\times
10^{-12}s$. The kinematic limit for this decay
is $\omega_{max}=1.284$. }
\label{dgdwf2bs}
\end{figure}

\clearpage
\newpage

\begin{figure}[p]
\epsfxsize = 5.4in \epsfbox{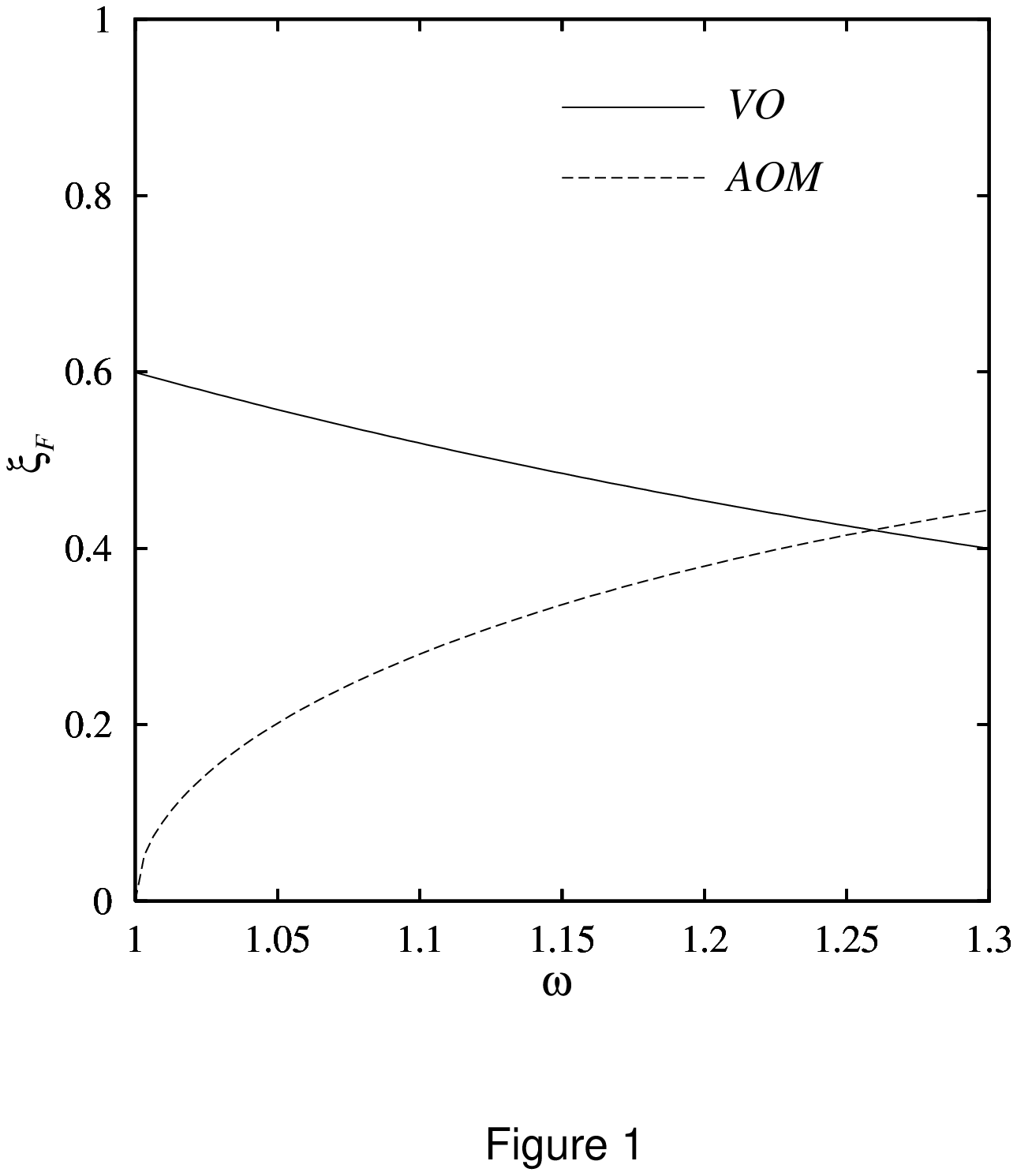}
\end{figure}

\begin{figure}[p]
\epsfxsize = 5.4in \epsfbox{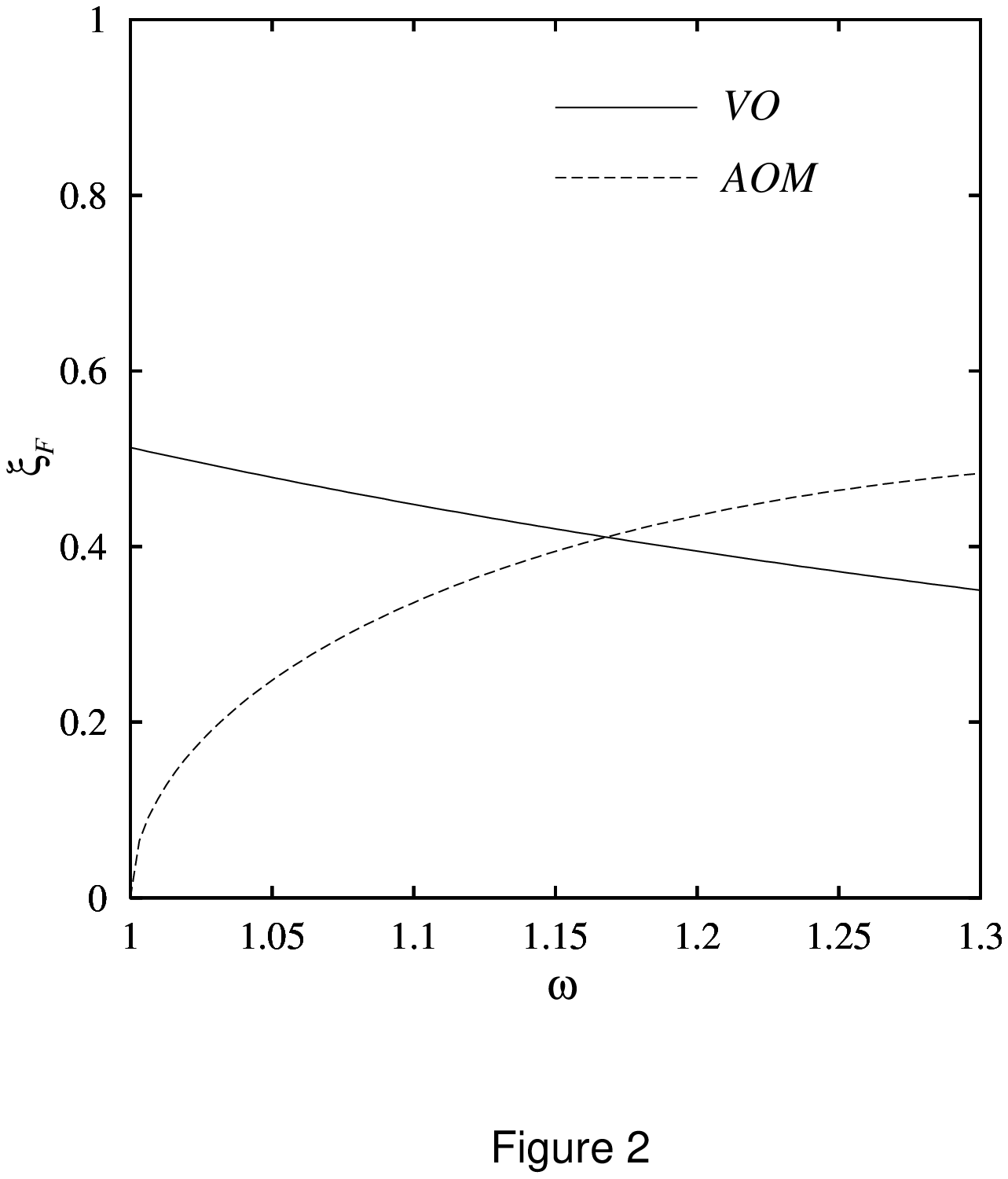}
\end{figure}

\begin{figure}[p]
\epsfxsize = 5.4in \epsfbox{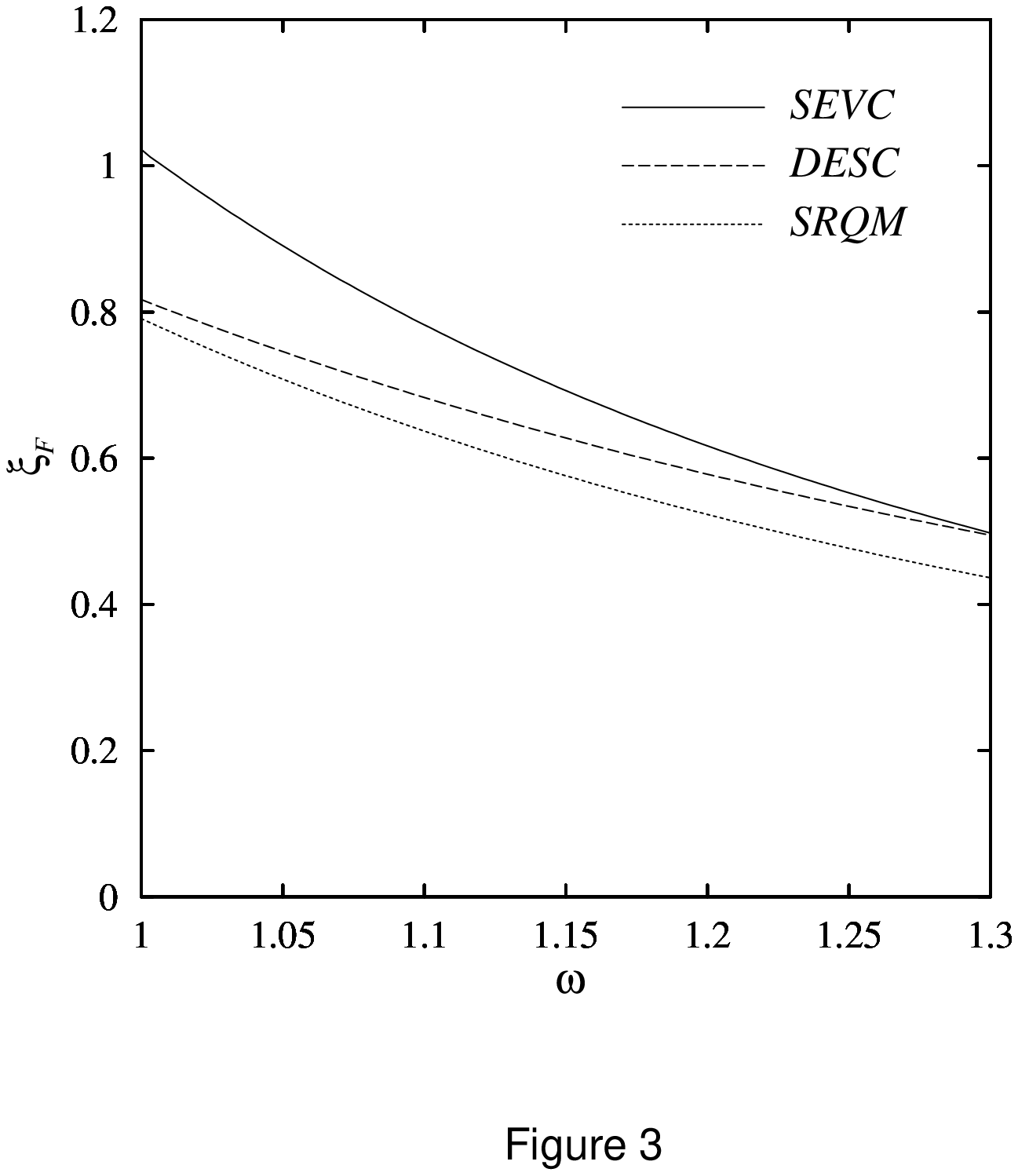}
\end{figure}

\begin{figure}[p]
\epsfxsize = 5.4in \epsfbox{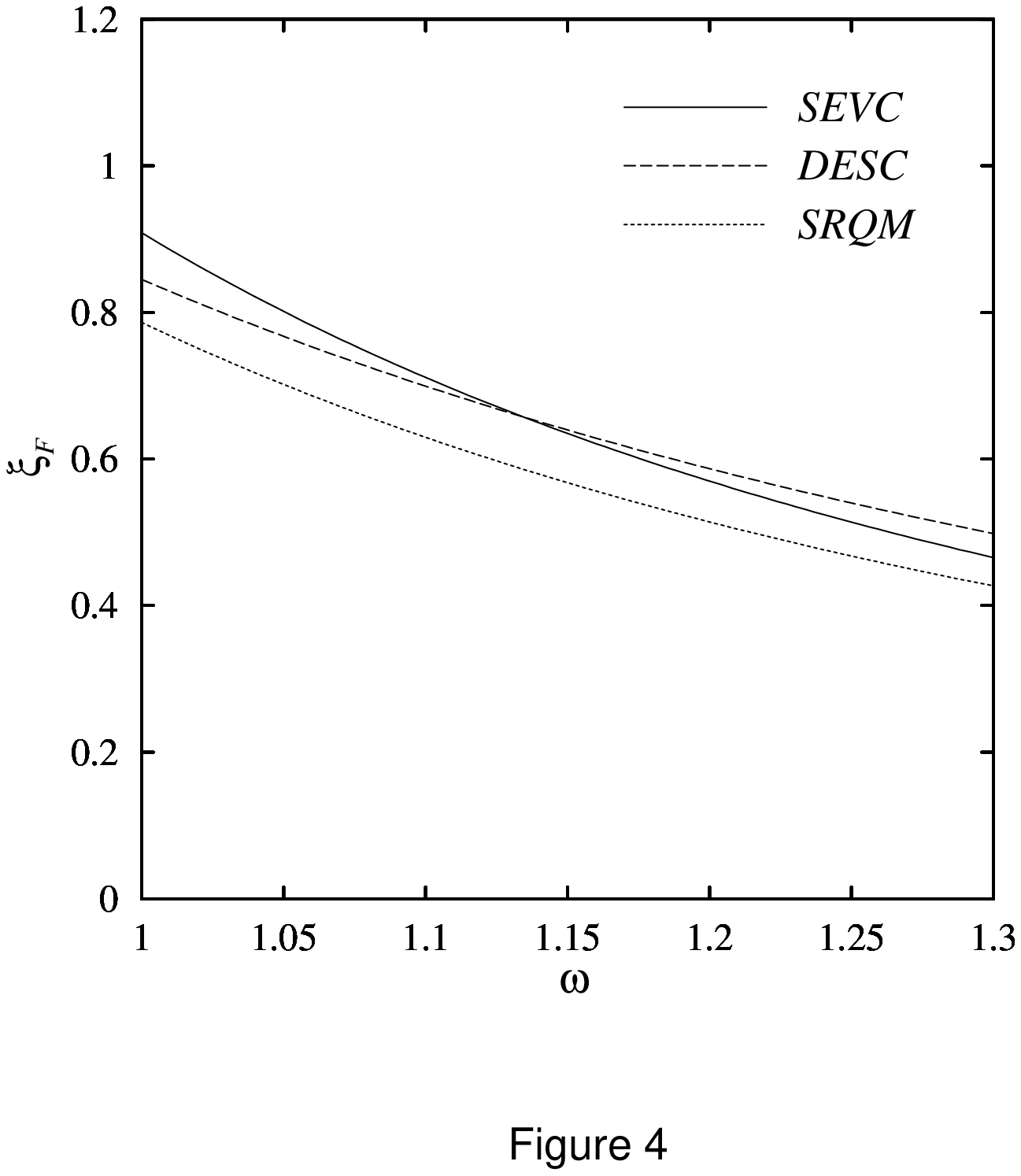}
\end{figure}

\begin{figure}[p]
\epsfxsize = 5.4in \epsfbox{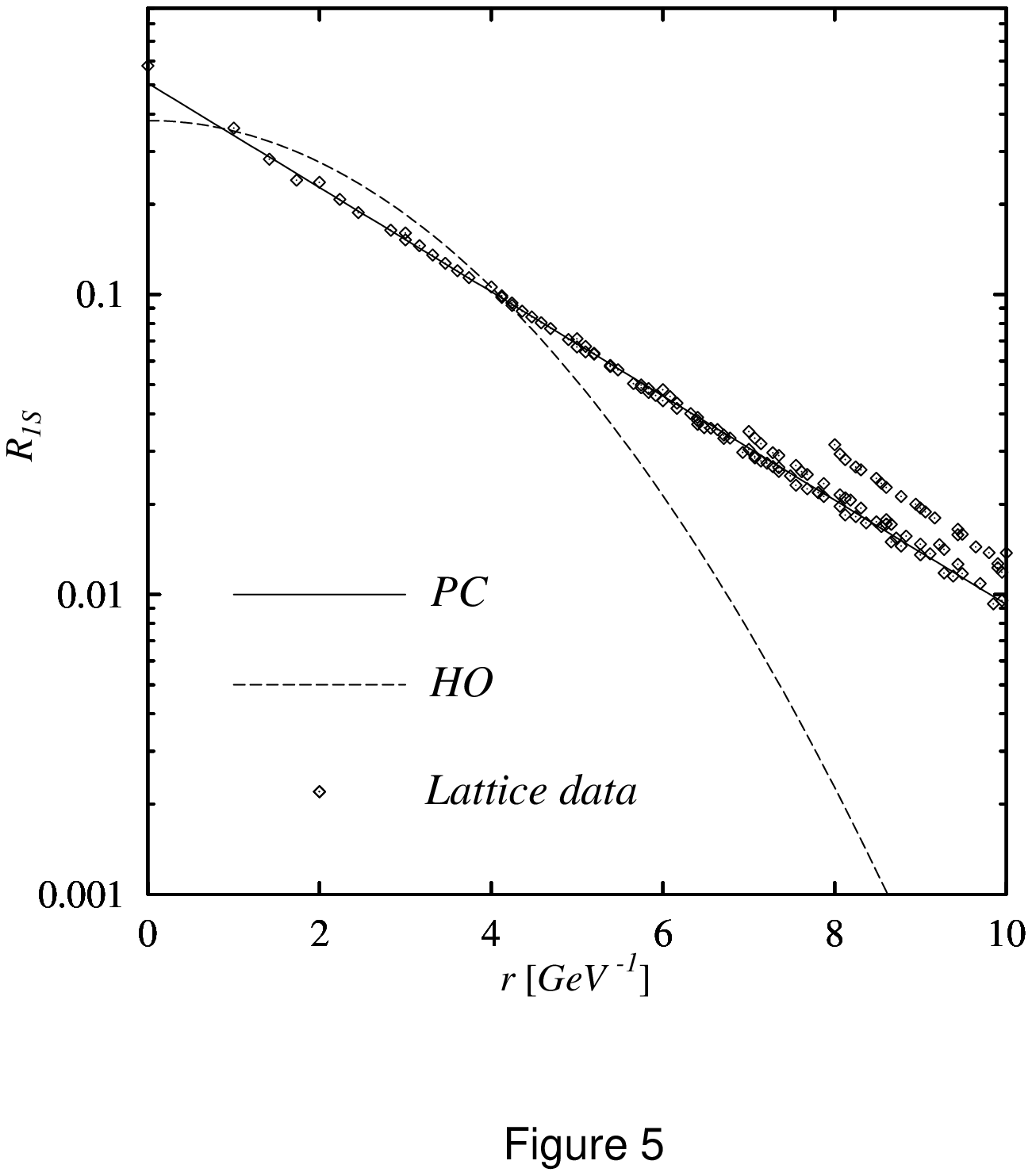}
\end{figure}

\begin{figure}[p]
\epsfxsize = 5.4in \epsfbox{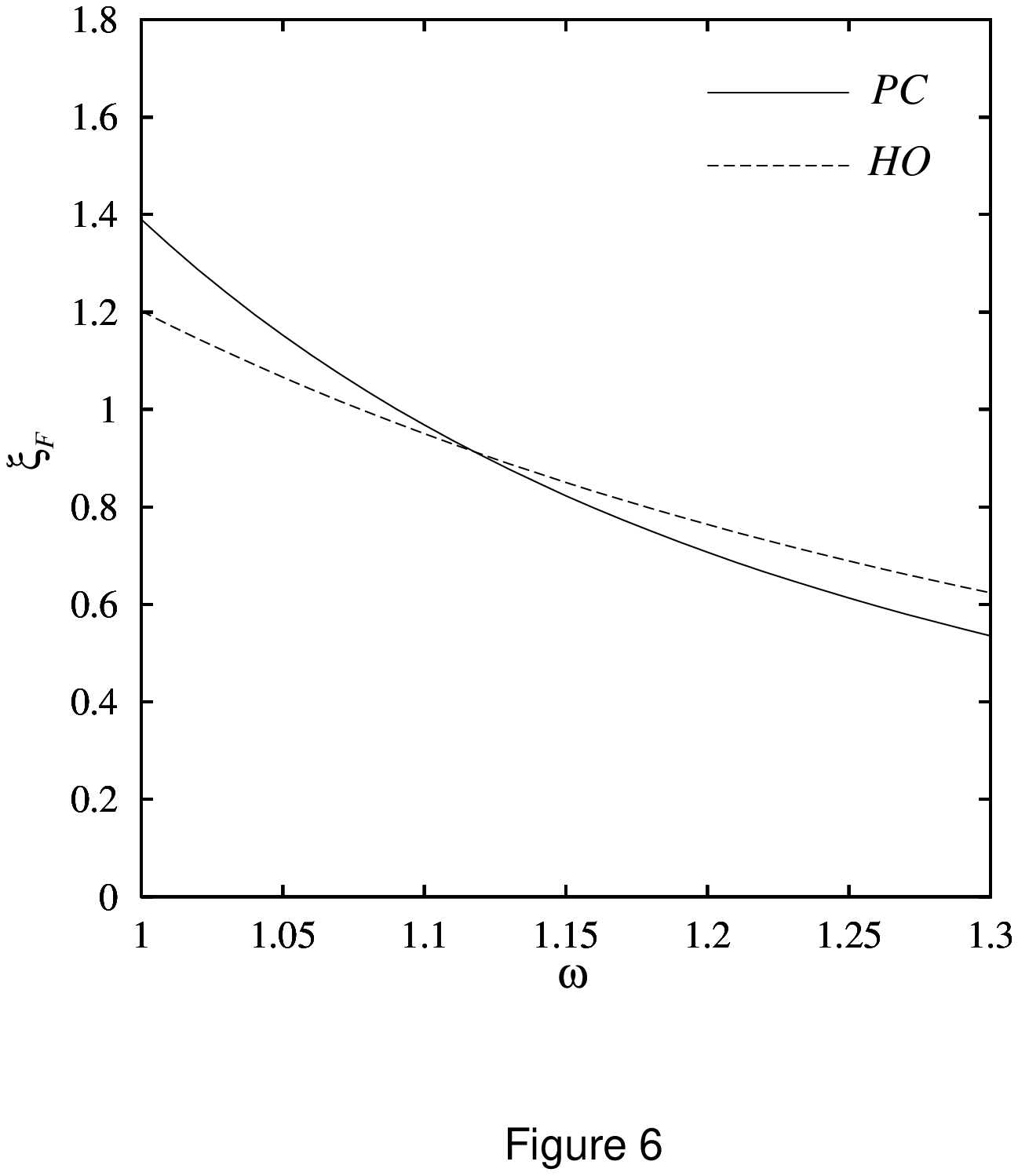}
\end{figure}

\begin{figure}[p]
\epsfxsize = 5.4in \epsfbox{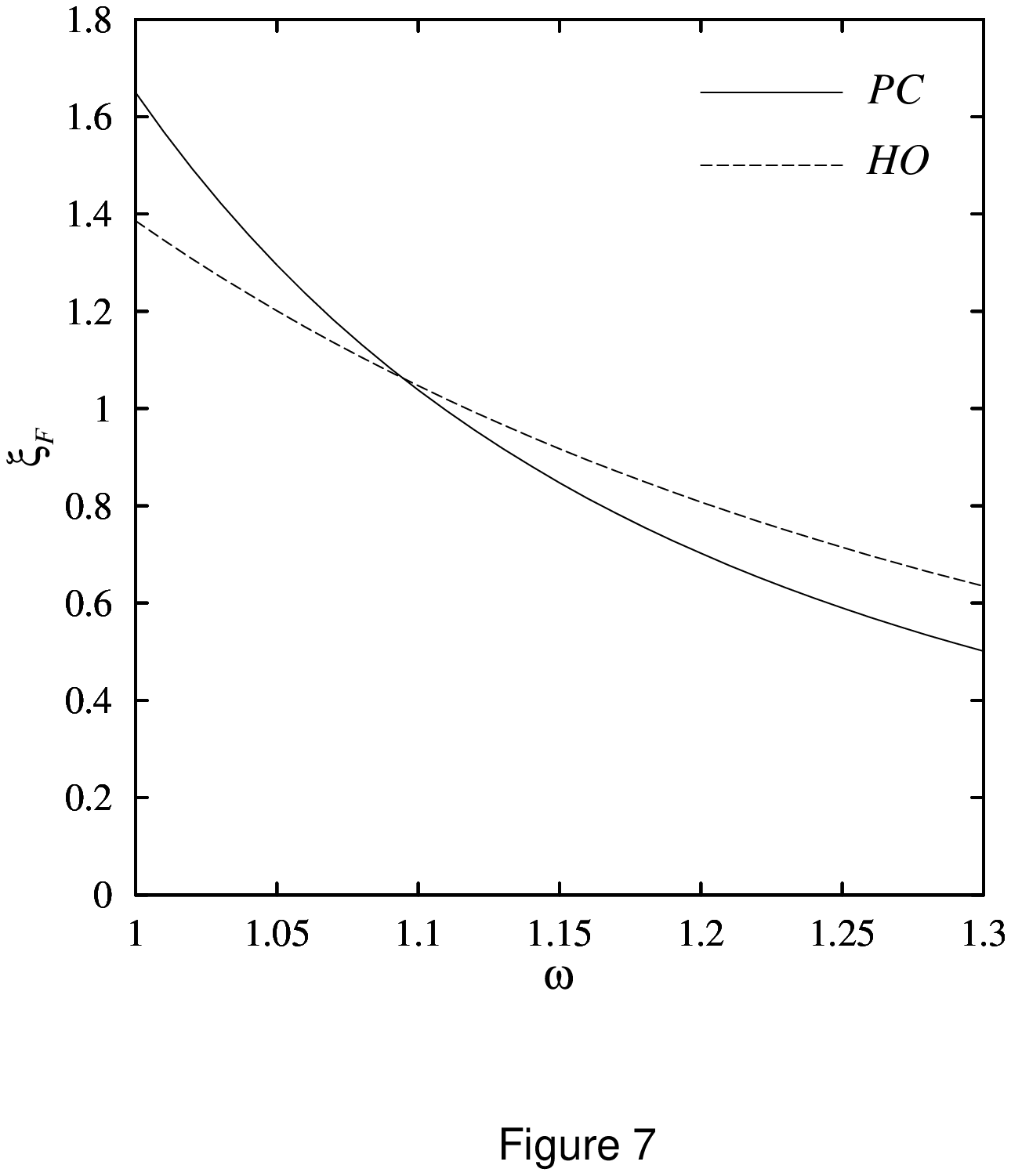}
\end{figure}

\begin{figure}[p]
\epsfxsize = 5.4in \epsfbox{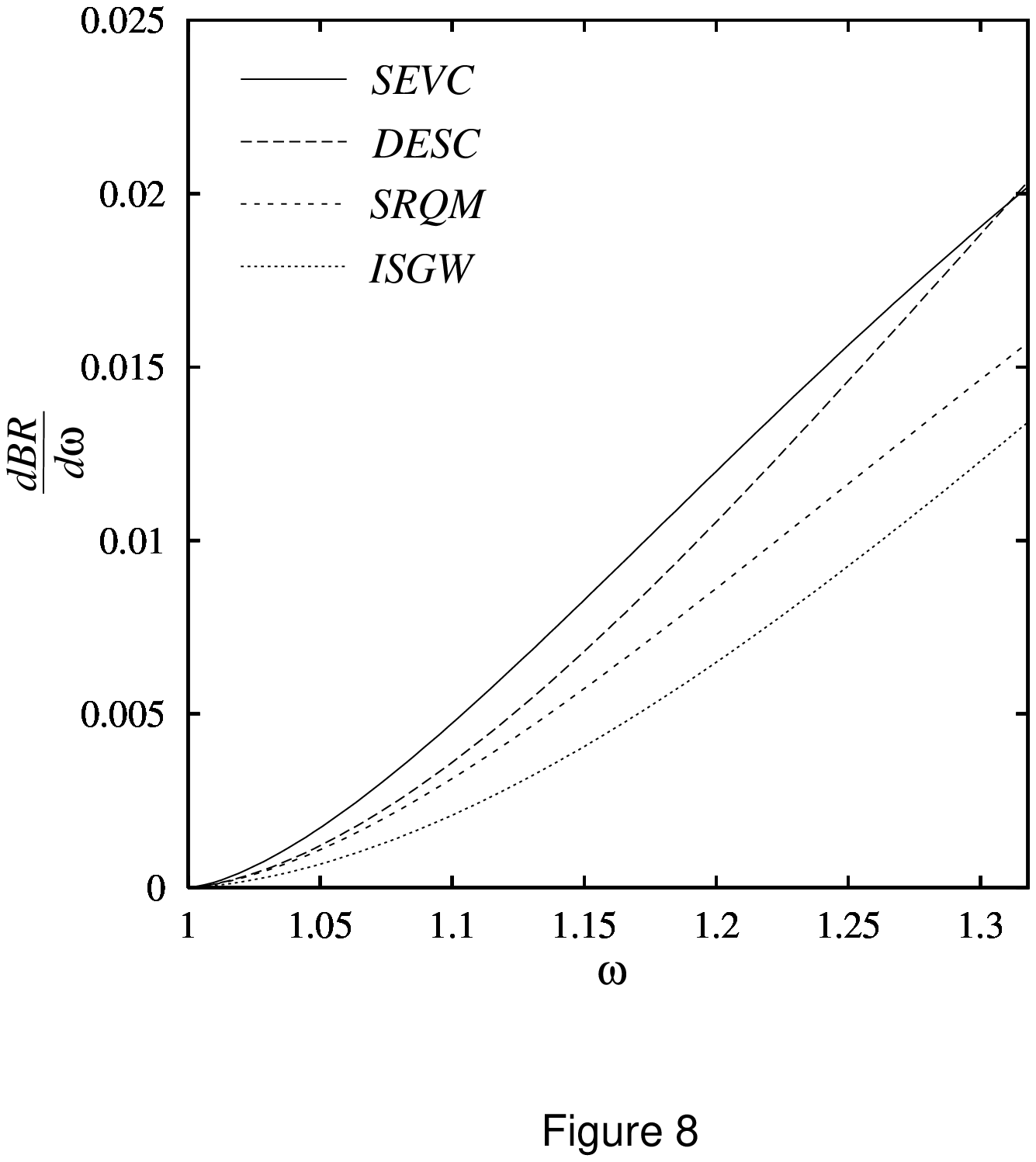}
\end{figure}

\begin{figure}[p]
\epsfxsize = 5.4in \epsfbox{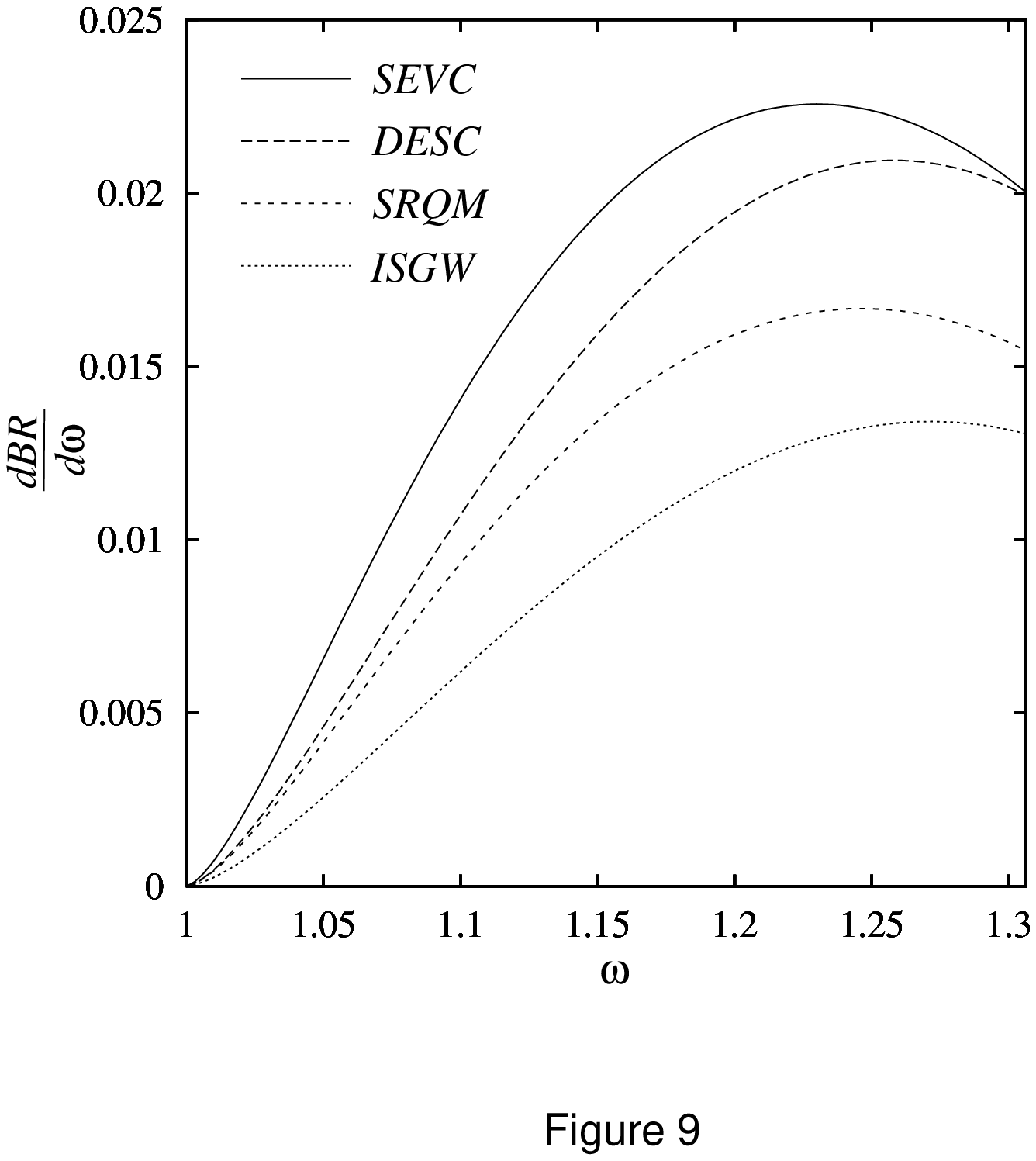}
\end{figure}

\begin{figure}[p]
\epsfxsize = 5.4in \epsfbox{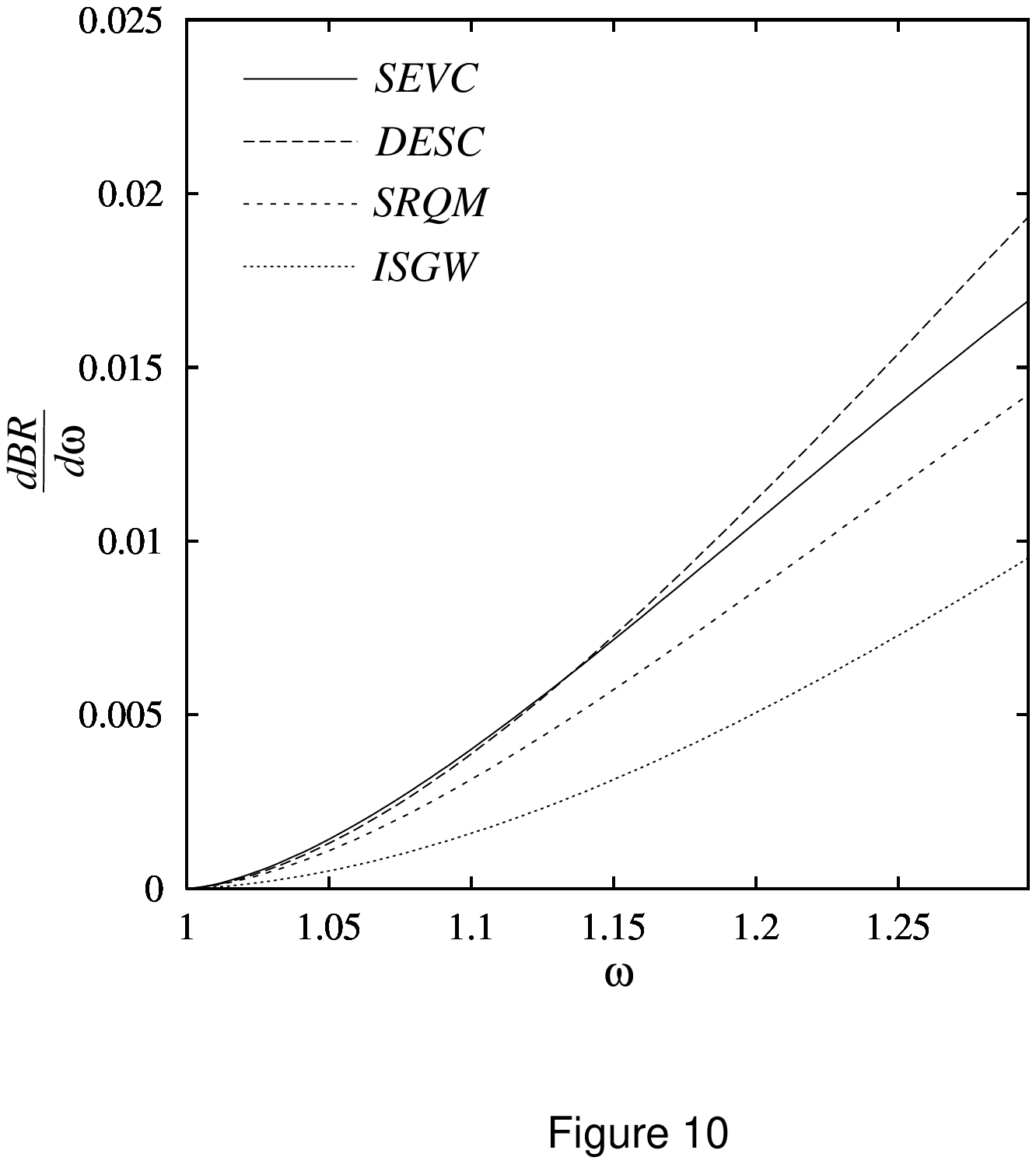}
\end{figure}

\begin{figure}[p]
\epsfxsize = 5.4in \epsfbox{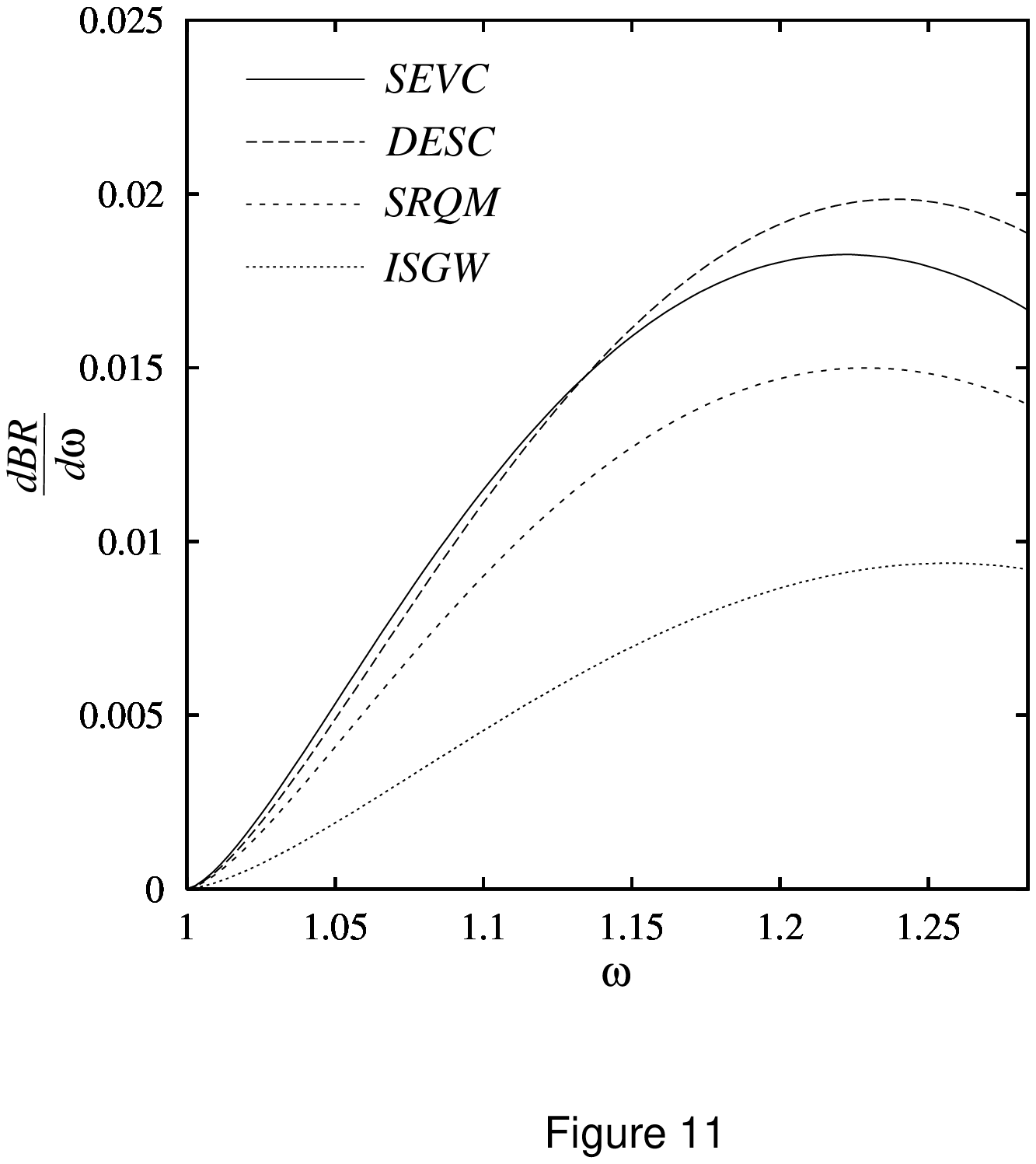}
\end{figure}

\end{document}